\documentclass[lettersize,journal]{IEEEtran}
\usepackage{amsmath,amsfonts}
\usepackage{algorithmic}
\usepackage{algorithm}
\usepackage{array}
\usepackage[caption=false,font=footnotesize]{subfig}
\usepackage{textcomp}
\usepackage{stfloats}
\usepackage{url}
\usepackage{verbatim}
\usepackage{graphicx}
\usepackage{cite}
\usepackage{color}
\usepackage{amsmath}
\usepackage{multirow}
\usepackage{tabularx}
\usepackage{ragged2e}
\usepackage[normalem]{ulem}
\usepackage{steinmetz}
\usepackage{titlesec}
\titlespacing*{\subsection}
{0pt}    
{0.5\baselineskip} 
{0.2\baselineskip}  
\begin{document}
\title{An Inductively Assisted $N$:1 Switched-Capacitor Ladder Converter}

\author{\IEEEauthorblockN{Sandeep Reddy Kukunuru, M Mahmudul Hasan Sajeeb, Rohit Reddy Karnaty, and Loai G. Salem}
\thanks{The authors are with the Department of Electrical and Computer Engineering, University of California at Santa Barbara, Goleta, CA 93106 USA (e-mail: sandeepreddy@ucsb.edu; msajeeb@ucsb.edu; rkarnaty@ucsb.edu;  salem@ucsb.edu).}
}



\maketitle

\begin{abstract}
This paper proposes an inductively assisted $N$:1 switched-capacitor ladder buck (D-SCLB) converter that replaces the output-side switch in the bottom half-bridge of an $N$:1 SCLB with a small inductor placed in parallel with the ladder capacitors during the demagnetization phase. The flying capacitors share the load current delivery, lowering the average inductor current to $(1{-}M)I_{out}$ and enabling continuous regulation with $M{=}D/(N{-}1{+}D)$. A closed-form output-impedance model, valid for arbitrary $N$ under both optimal and uniform switch sizing, is derived and compared against SC, SCLB, FCMC, and DPSC topologies. For a 4:1 48\,V-to-PoL design, the D-SCLB reduces the average inductor current by up to 20\% and improves the equivalent output impedance by up to 30\% relative to the SCLB. Two hardware prototypes on a low-cost two-layer PCB achieve a peak efficiency of 96.2\%, with the D-SCLB using 49\% smaller inductor volume than the SCLB at comparable efficiency. Model predictions agree with measurements to within 1.5\% under heavy load.
\end{abstract}


\section{Introduction}

\begin{figure}[!tb]
\vspace*{0cm} 
\centering
\includegraphics[width=3.4in]{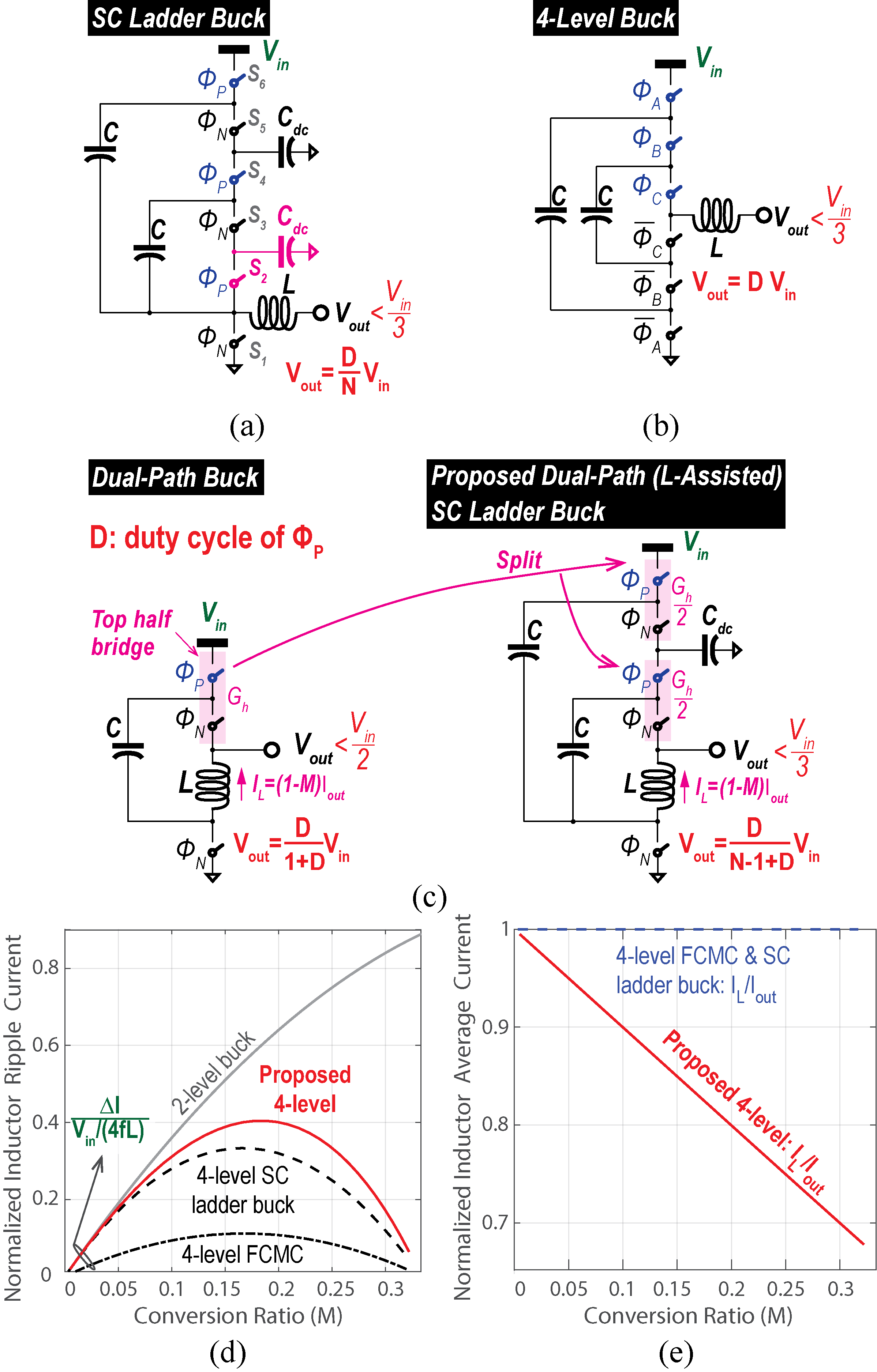}
\vspace*{0cm} 
\caption{\textcolor{black}{Prior (a) SC ladder and (b) flying capacitor multilevel buck converters. (c) Proposed dual-path SC ladder buck converter. Comparison of the (d) inductor ripple current and (e) average current in a 4-level SC ladder buck, 4-level FCMC, and the proposed topology.}}
\label{fig.proposed_dsclb}
\end{figure}

\begin{figure*}[!tb]%
 \vspace*{0 cm} 
 \centering
 \subfloat[]{\includegraphics[width=0.32\textwidth]{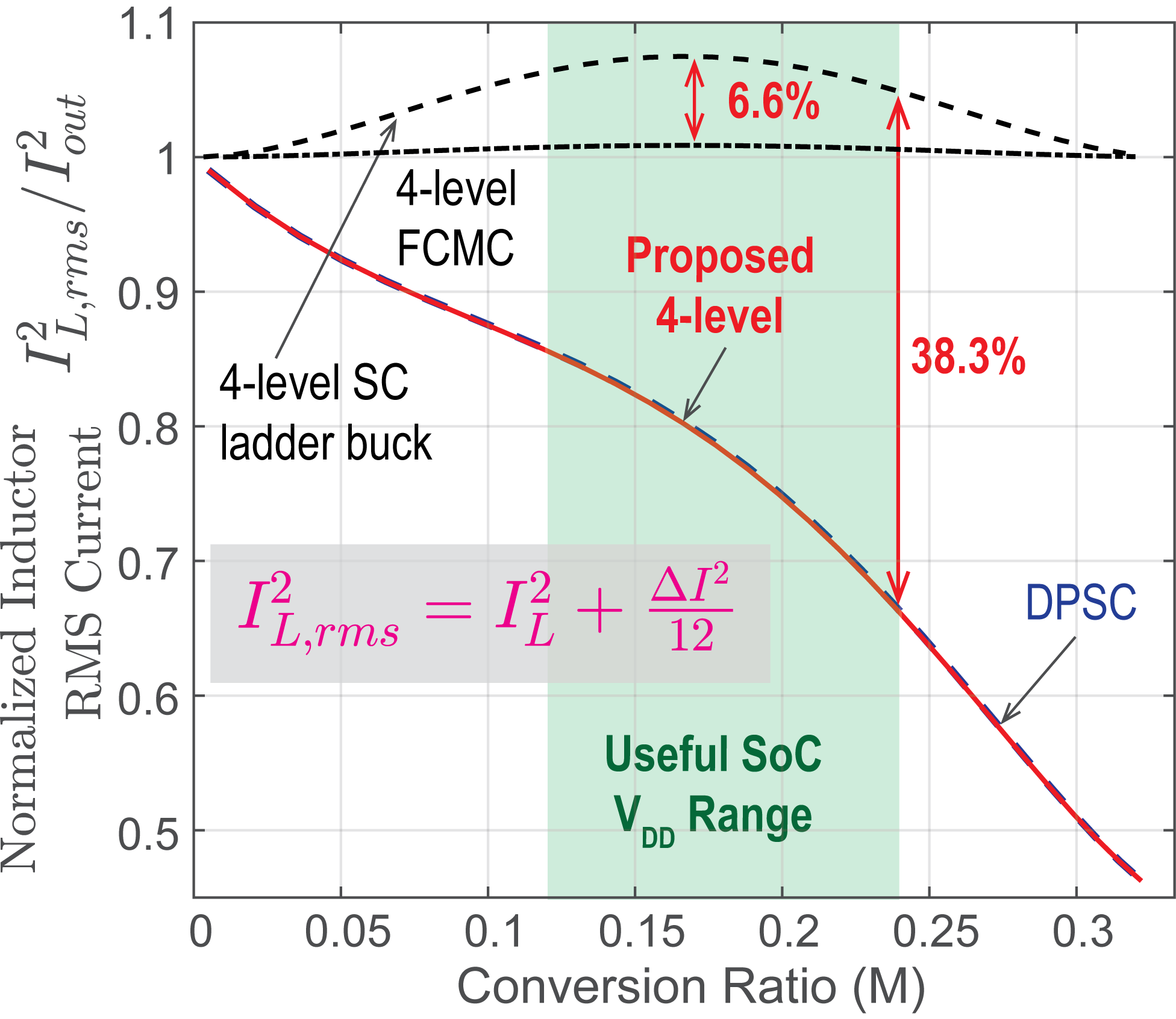}\label{fig.sim_res_a}}%
 \hspace{0.1cm}
 \subfloat[]{\includegraphics[width=0.304\textwidth]{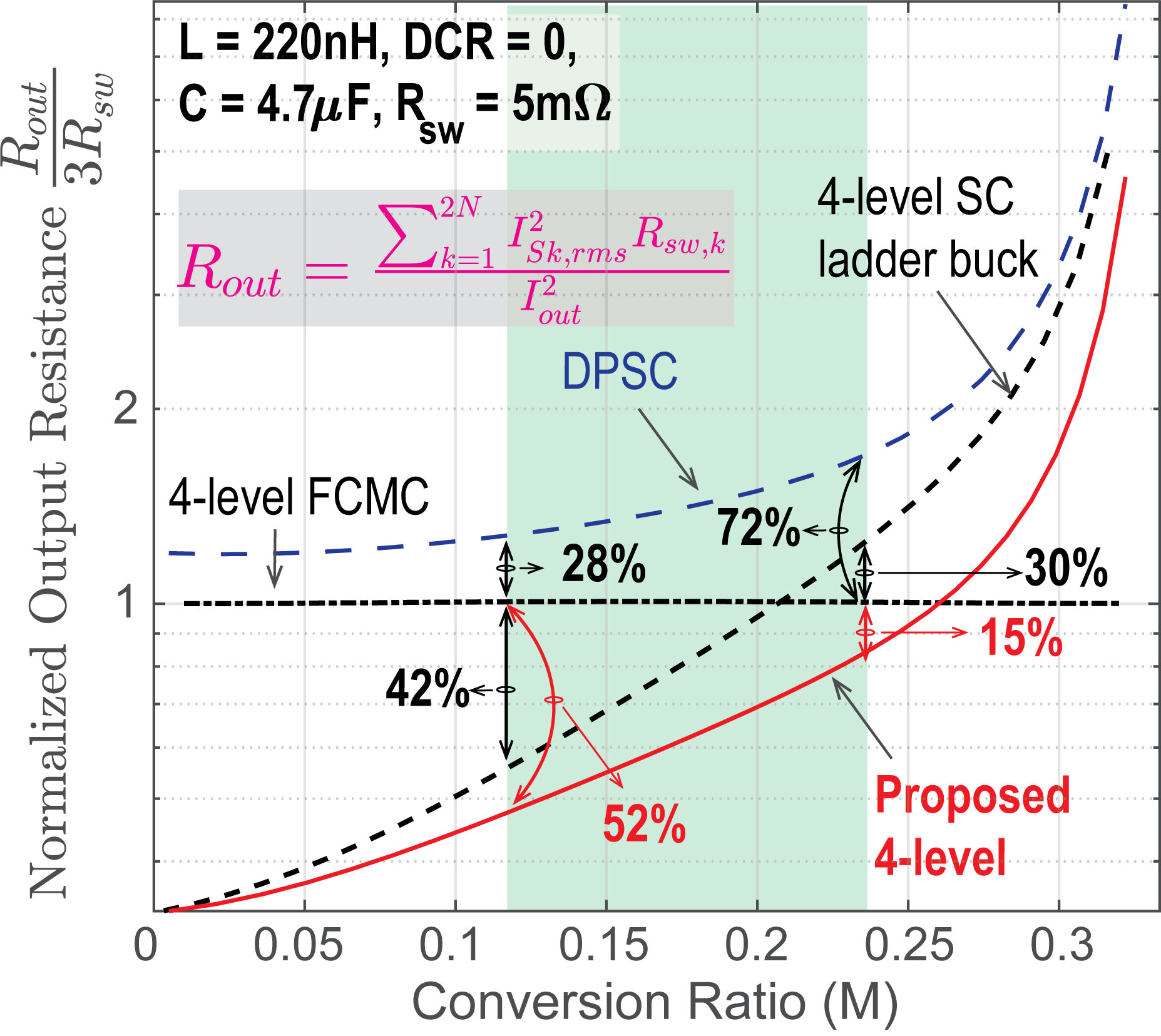}\label{fig.sim_res_b}}%
 \hspace{0.1cm}
 \subfloat[]{\includegraphics[width=0.32\textwidth]{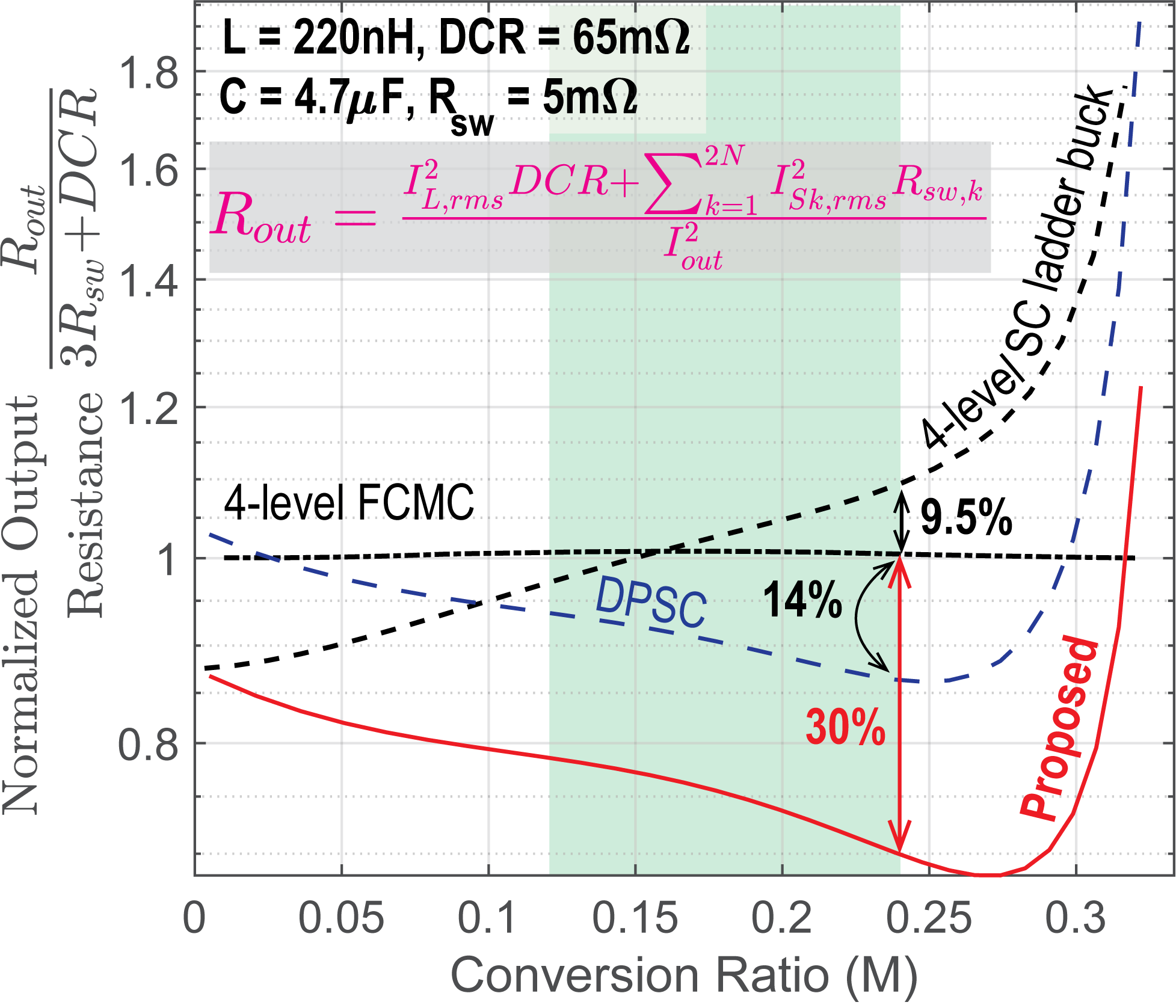}\label{fig.sim_res_c}}%
  \caption{\textcolor{black}{(a) Comparison of the inductor RMS current squared of a 4-level SC ladder buck, 4-level FCMC, DPSC, and the proposed topology. Comparison of the normalized equivalent output resistance of the four converters while employing the same total conductance (switch area) when DCR is (b) 0 and (c) 65m$\Omega$. }}%
 \label{fig.sim_res}%
  \vspace*{0 cm} 
\end{figure*}

The growing demand for USB-powered consumer electronics requires DC-DC voltage regulators that can provide high N:1 voltage conversion without sacrificing efficiency. \textcolor{black}{Furthermore, most high-performance computing platforms aim to operate their SoCs from an N$\times$ higher supply voltage, $V_{in}$, to lower the input current by a factor of N and ease otherwise large $I$\textsuperscript{2}${R}$ losses in power distribution.} Switched-capacitor (SC) ladder buck (SCLB) converters \cite{b1,b2} have emerged as a prime candidate for high N:1 voltage conversion, where the reduced switch rated-voltage, $V_{in}$/N, allows the power converter to employ the same thin-oxide transistors used to implement the powered SoC digital processing \textcolor{black}{to realize the power switches} \textcolor{black}{(Fig. \ref{fig.proposed_dsclb}a)}. (N+1)-level flying capacitor multilevel converters (FCMC) \cite{b3} offer the same advantage (Fig. \ref{fig.proposed_dsclb}b), and furthermore, they exhibit an N$\times$ lower inductor peak-to-peak current ripple ($\Delta I$), as shown in Fig. \ref{fig.proposed_dsclb}d, where the frequency of voltage swing across L is N$\times$ that of a SCLB. As a result, FCMCs feature a smaller overall inductor RMS current ($I_{L,rms}$) compared to an (N+1)-level SCLB (Fig. \ref{fig.sim_res}a). Despite the higher $I_{L,rms}$, a SCLB still achieves a lower equivalent output resistance ($R_{out}$), e.g., $<$42\% for a 4-level converter (Fig. \ref{fig.sim_res}b), in contrast to a FCMC across the useful SoC $V_{DD}$ range. The reason behind that is, unlike a FCMC, a SCLB does not employ stacked low-side switches, while the pull-up switches in the N half bridges of a SCLB can be considered effectively connected in parallel in the AC sense \cite{b4}. The ladder topology embedded within a SCLB splits the inductor current, $I_{L}$, optimally among the ladder switches such that the overall $R_{out}$ is minimized, unlike FCMCs that mainly aim to minimize $\Delta$I. 

In fact, the SC ladder topology meets the fundamental limit for minimum switch $V$-$A$ stress applicable to all DC-DC converters for a given conversion ratio and under identical load currents, $I_{out}$ \cite{b4}.Unfortunately, when the inductor \textcolor{black}{DCR} is large compared to the switch on-resistance, $R_{sw}$, in designs of high power density, the higher $I_{L,rms}$ increases $R_{out}$ of a SCLB above that of a FCMC, e.g., by 9.5\% across the useful SoC $V_{DD}$ range (Fig. \ref{fig.sim_res}c). \textcolor{black}{Unlike SCLB and FCML converters, inductively-assisted (dual-path) SC converters (LSC) \cite{b5} can reduce the average component of the inductor current ($I_{L}$) below $I_{out}$, in addition to lowering the inductor ripple current component, wherein the embedded SC supplies the remaining amount of $I_{out}$ while reducing the voltage swing across L. Unfortunately, the parent SC topology of existing LSC converters is typically a series-parallel \cite{b6} or a binary \cite{b13} topology, and hence, existing LSC converters\textcolor{black}{\cite{b14}} do not take advantage of the superior switch performance of a SC ladder topology to realize efficient steep DC-DC conversion.}


\section{Proposed Dual-Path Switched-Capacitor Ladder Buck Converter}
\label{prop_dsclb}
To overcome the high \textcolor{black}{DCR} loss problem in a SCLB, we propose to place the inductor in parallel with the ladder capacitors during the inductor demagnetization phase to lower the inductor average current component ($I_{L}$) below load current ($I_{out}$). The proposed inductively assisted SC ladder or dual-path SCLB (D-SCLB) in Fig. \ref{fig.proposed_dsclb}c replaces the $V_{out}$-side switch in the bottom half bridge of a SC ladder DC-DC converter with a small inductor to enable continuous voltage conversion below the native SC ladder discrete ratio, 1/N (conversion ratio M = D/[N-1+D]), \textcolor{black}{where D is the fraction of the cycle time during which $\Phi_{P}$ switches are on}. 
The (N-1) capacitors in the flying or fixed ladder can be connected in series rather than the parallel connection shown in Fig. 1c (right). However, placing the capacitors of a ladder in parallel with their bottom plates tied together minimizes the slow-switching-limit (SSL) C$\Delta ^2$-based loss for a given capacitance\textcolor{black}{\cite{b15}} and permits soft-charging of the capacitors.\\
The proposed D-SCLB can be viewed as if the top half bridge in a dual-path buck (Fig. \ref{fig.proposed_dsclb}c, left) \cite{b5,b6} and its coupling capacitance are split into N-1 identical half bridges that are stacked on top of each other. The flying ladder automatically balances the voltage across the stacked half bridges at ($V_{in}$-$V_{out}$)/(N-1). In contrast to a SCLB, both L and C in parallel supply current to $V_{out}$, while $\phi_{N}$=1, which lowers the inductor average current, $I_{L}$, below $I_{out}$, i.e., $I_{L}$=(1-M)$I_{out}$, as shown in Fig. \ref{fig.proposed_dsclb}e. Therefore, while $\Delta$$I$ of the proposed D-SCLB is higher than that of a SCLB and FCMC, the inductor overall squared RMS current due to both its average and ripple components ($I_{L,rms}$\textsuperscript{2} = $I_{L}$\textsuperscript{2} + $\Delta$$I$\textsuperscript{2}/12) in the D-SCLB is up to 50\% (at M=0.3) lower than the other two converters (Fig. \ref{fig.sim_res}a). Note that the \textcolor{black}{DCR} loss due to $\Delta$$I$ is 12$\times$ smaller than that due to $I_{L}$. While a dual-path series-capacitor (DPSC) converter \cite{b7} features identical $I_{L}$ reduction advantage (Fig. \ref{fig.sim_res}a), the number of switches in a DPSC and their voltage rating are larger than that of a D-SCLB under the same M. Therefore, using the same switch area and identical inductors, the proposed D-SCLB achieves $R_{out}$ improvements over a SCLB and DPSC of 45\% and 87\% (\textcolor{black}{DCR}=0; Fig. \ref{fig.sim_res}b), and 39.5\% and 16\% (\textcolor{black}{DCR}=65m$\Omega$; Fig. \ref{fig.sim_res}c) under the same $V_{in}$, $I_{out}$, and $f$.\\

\section{Steady-State Comparative Analysis}

\begin{figure}[!tb]
  \centering
  \includegraphics[width=3.4in]{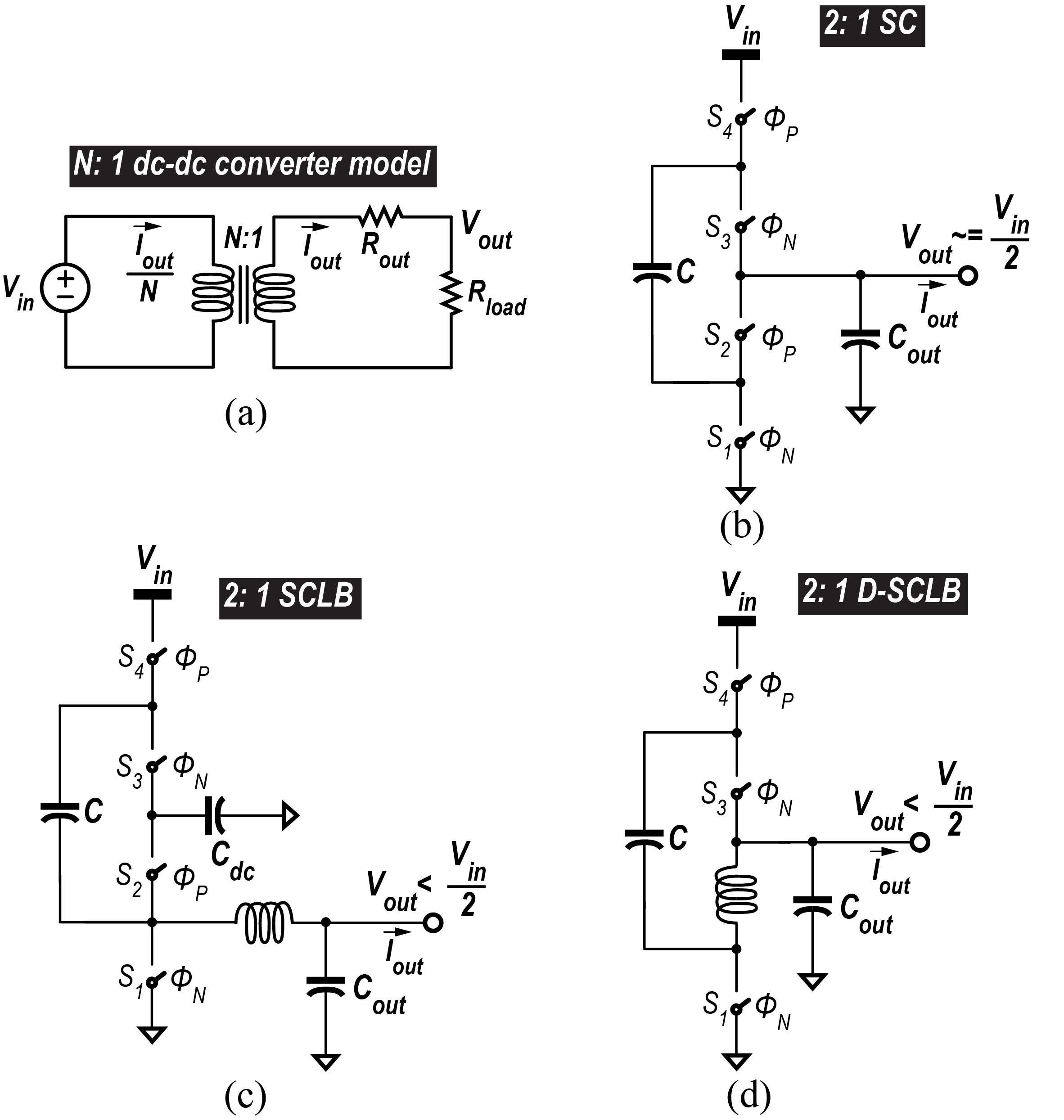}
  \caption{(a) Steady-state equivalent model of an $N$:1 dc–dc converter. Schematic diagrams of (b) a 2:1 SC converter, (c) a 2:1 SCLB converter, and (d) a 2:1 D-SCLB converter.}
  \label{fig.2x1scmodel}
\end{figure}

A dc–dc converter with an $N$:1 conversion ratio can be abstractly represented by an ideal $N$:1 transformer to capture its voltage-scaling behavior, even though it operates on dc signals, as shown in Fig.~\ref{fig.2x1scmodel}(a).

\subsection{2:1 Converter Output Impedance}

\begin{table}[!th]
    \centering
    \caption{Normalized Switch RMS Current}
    \renewcommand{\arraystretch}{1.5}
    \setlength{\tabcolsep}{14pt}
    \begin{tabular}{c|c|c|c}
        \hline\hline
        \multirow{2}{*}{\textbf{Switch}} & \multicolumn{3}{c}{\textbf{$I_{RMS,i}/I_{out}$}} \\
        \cline{2-4}
        & \textbf{2:1 SC} & \textbf{2:1 SCLB} & \textbf{2:1 D-SCLB} \\
        \hline
        $S_1$ & $\frac{1}{2\sqrt{1-D}}$ & $\frac{2-D}{2\sqrt{1-D}}$ & $\frac{1}{(1+D)\sqrt{1-D}}$ \\
        $S_2$ & $\frac{1}{2\sqrt{D}}$ & $\frac{\sqrt{D}}{2}$ & -- \\
        $S_3$ & $\frac{1}{2\sqrt{1-D}}$ & $\frac{D}{2\sqrt{1-D}}$ & $\frac{D}{(1+D)\sqrt{1-D}}$ \\
        $S_4$ & $\frac{1}{2\sqrt{D}}$ & $\frac{\sqrt{D}}{2}$ & $\frac{\sqrt{D}}{1+D}$ \\
        \hline\hline
    \end{tabular}
    \label{tab:normalized_rms_current_2x1}
\end{table}
The 2:1 SC, 2:1 SCLB, and 2:1 D-SCLB converters are shown in Fig.~\ref{fig.2x1scmodel}(b)–(d), respectively, and their normalized switch RMS currents, $I_{\mathrm{RMS},i}/I_{\mathrm{out}}$, are listed in Table~\ref{tab:normalized_rms_current_2x1}, where $D$ is the duty cycle of the clock phase $\phi_{P}$. To compare the three topologies, the total switch area (conductance, $G$) is assumed identical, and each switch is sized for minimum conduction loss according to
{%
\setlength{\abovedisplayskip}{0.5\baselineskip}%
\setlength{\belowdisplayskip}{0.5\baselineskip}%
\begin{equation}
G_{i} = \frac{I_{RMS,i}}{\Sigma I_{RMS}}G,
\label{eq:rsizingfund}
\end{equation}}%
where $G_{i}$ is the conductance of the switch carrying $I_{RMS,i}$ and $\Sigma I_{RMS}$ is the sum of the RMS currents of all switches. Although both the RMS currents and the optimal switch sizes depend on $D$, the nominal point $D = 0.5$ is adopted for this analysis. The resulting optimal relative switch sizes, assuming equal total conductance, are summarized in Table~\ref{tab:optimal_relsizing_2x1}. 

\begin{table}[!th]
    \centering
    \caption{Optimal Switch Relative Sizing for a Fixed Total Conductance, $G$, at $D = 0.5$}
    \renewcommand{\arraystretch}{1.5}
    \setlength{\tabcolsep}{14pt}
    \begin{tabular}{c|c|c|c}
        \hline\hline
        \multirow{2}{*}{\textbf{Switch}} & \multicolumn{3}{c}{\textbf{Converter Topology}} \\
        \cline{2-4}
        & \textbf{2:1 SC} & \textbf{2:1 SCLB} & \textbf{2:1 D-SCLB} \\
        \hline
        $S_1$ & $\frac{G}{4}$            & $\frac{G}{2}$     & $\frac{G}{2}$ \\
        $S_2$ & $\frac{G}{4}$            & $\frac{G}{6}$   & --                \\
        $S_3$ & $\frac{G}{4}$            & $\frac{G}{6}$   & $\frac{G}{4}$          \\
        $S_4$ & $\frac{G}{4}$            & $\frac{G}{6}$   & $\frac{G}{4}$          \\
        \hline\hline
    \end{tabular}
    \label{tab:optimal_relsizing_2x1}
\end{table}

The equivalent output impedance of a 2:1 switched-capacitor (SC) converter operating in the fast-switching limit (FSL) is given by
{%
\setlength{\abovedisplayskip}{0.5\baselineskip}%
\setlength{\belowdisplayskip}{0.5\baselineskip}%
\begin{equation}
R_{out} = \frac{1}{G}\times\frac{2}{D(1-D)},
\label{eq:rout2x1}
\end{equation}}%
\textcolor{black}{switching losses associated with the MOSFET capacitances and their intrinsic body-drain pn-junction diodes are not included in this model.}

The ideal conversion ratio, \textcolor{black}{\(M\,=\,V_{out}/V_{in}\)}, of the SCLB converter in Fig.~\ref{fig.2x1scmodel}(c) is given by
{%
\setlength{\abovedisplayskip}{0.5\baselineskip}%
\setlength{\belowdisplayskip}{0.5\baselineskip}%
\begin{equation}
M = \frac{D}{2},
\end{equation}}%
while the average inductor current, $I_L$, and the ripple current, $\Delta I$, are given by
{%
\setlength{\abovedisplayskip}{0.5\baselineskip}%
\setlength{\belowdisplayskip}{0.5\baselineskip}%
\begin{equation}
\begin{aligned}
I_L &= I_{out},\\
\Delta I &= \frac{D(1-D)}{2}\times\frac{V_{in}}{fL}.
\end{aligned}
\end{equation}}%
The equivalent output impedance of the 2:1 SCLB converter is given by
{%
\setlength{\abovedisplayskip}{0.5\baselineskip}%
\setlength{\belowdisplayskip}{0.5\baselineskip}%
\begin{equation}
\begin{aligned}
R_{out}
  &= r_{DCR}
  + \frac{1}{G}\times
    \frac{(2-D)(1+D)}{(1-D)}\\
  &\quad+ \frac{\Delta I^{2}}{12\,I_{out}^{2}}
      \biggl(
        r_{ESR}
        + \frac{2+D}{G}
      \biggr),
\end{aligned}
\label{eq:rout2x1sclb}
\end{equation}}%
where $r_{DCR}$ is the inductor’s dc winding resistance and $r_{ESR}$ is its effective series resistance at the switching frequency.

The ideal conversion ratio of the 2:1 D-SCLB shown in Fig.~\ref{fig.2x1scmodel}(d) is given by
{%
\setlength{\abovedisplayskip}{0.5\baselineskip}%
\setlength{\belowdisplayskip}{0.5\baselineskip}%
\begin{equation}
M = \frac{D}{1 + D}.
\end{equation}}%
The average and ripple currents of the inductor are given by
{%
\setlength{\abovedisplayskip}{0.5\baselineskip}%
\setlength{\belowdisplayskip}{0.5\baselineskip}%
\begin{equation}
\begin{aligned}
I_L &= (1 - M)I_{out}, \\
\Delta I &= \frac{D(1 - D)}{1 + D}\times\frac{V_{in}}{fL}.
\end{aligned}
\end{equation}}%
And the equivalent output impedance of the 2:1 D-SCLB converter is given by
{%
\setlength{\abovedisplayskip}{0.5\baselineskip}%
\setlength{\belowdisplayskip}{0.5\baselineskip}%
\begin{equation}
\begin{aligned}
R_{out} &= (1 - M)^2 \left(r_{DCR} + \frac{2}{G}\left(\frac{1 + 2D}{1 - D}\right)\right) \\
        &+ \frac{\Delta I^2}{12 I_{out}^2}\left(r_{ESR} + \frac{2(1 + D)}{G}\right).
\end{aligned}
\label{eq:rout2x1dsclb}
\end{equation}}

\begin{figure}[!tb]
\centering
\includegraphics[width=2.4in]{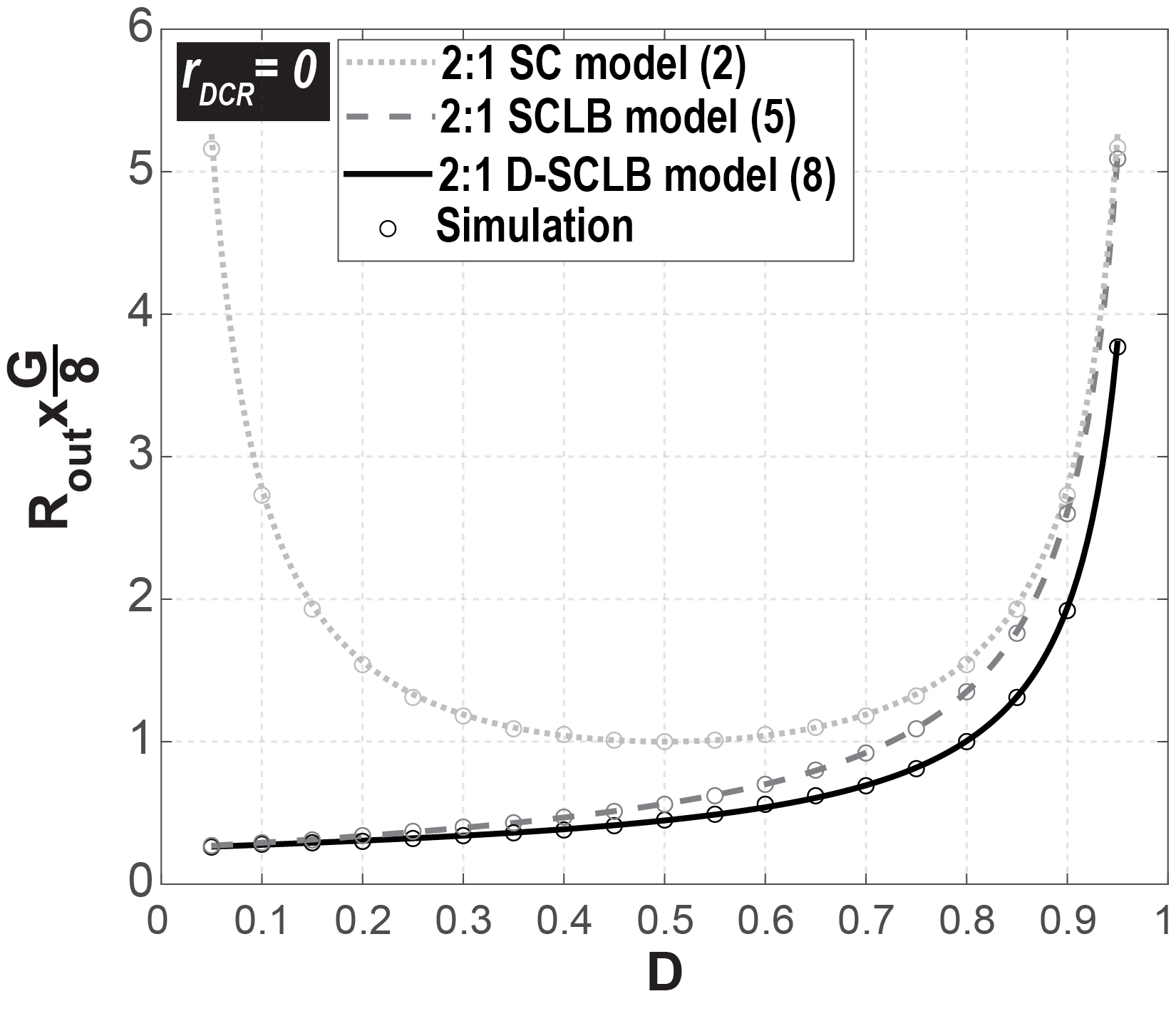}
\caption{Equivalent output impedance of the three converters versus duty cycle, $D$, with the inductor DCR and ESR equal to zero.}
\label{fig.2x1routcomp}
\end{figure}
Fig.~\ref{fig.2x1routcomp} illustrates the normalized equivalent output impedance of the three 2:1 converters as the duty cycle, $D$, is varied from 0 to 1, with $r_{DCR}$ and $r_{ESR}$ set to zero for a fair comparison with the 2:1 SC converter. The equivalent output impedance is normalized to the equivalent output impedance value of the 2:1 SC at $D$\,=\,0.5, which is equal to $\frac{8}{G}$. As shown, the output impedance of the 2:1 SC converter is optimal at $D$\,=\,0.5 and rises significantly as $D$ approaches either 0 or 1. Consequently, SC converters typically operate at $D$\,=\,0.5 to minimize $R_{out}$ for a given total conductance, $G$. In contrast, the 2:1 SCLB and 2:1 D-SCLB converters achieve their minimum $R_{out}$ at $D$\,=\,0 and increase hyperbolically as $D$ nears unity, due to the asymptotic behavior caused by the $(1-D)$ denominator. Since the 2:1 D-SCLB converter uses one fewer switch compared to the other two converters, it maintains a superior $R_{out}$ across the entire duty-cycle range.

\begin{figure}[!tb]
\centering
\includegraphics[width=2.4in]{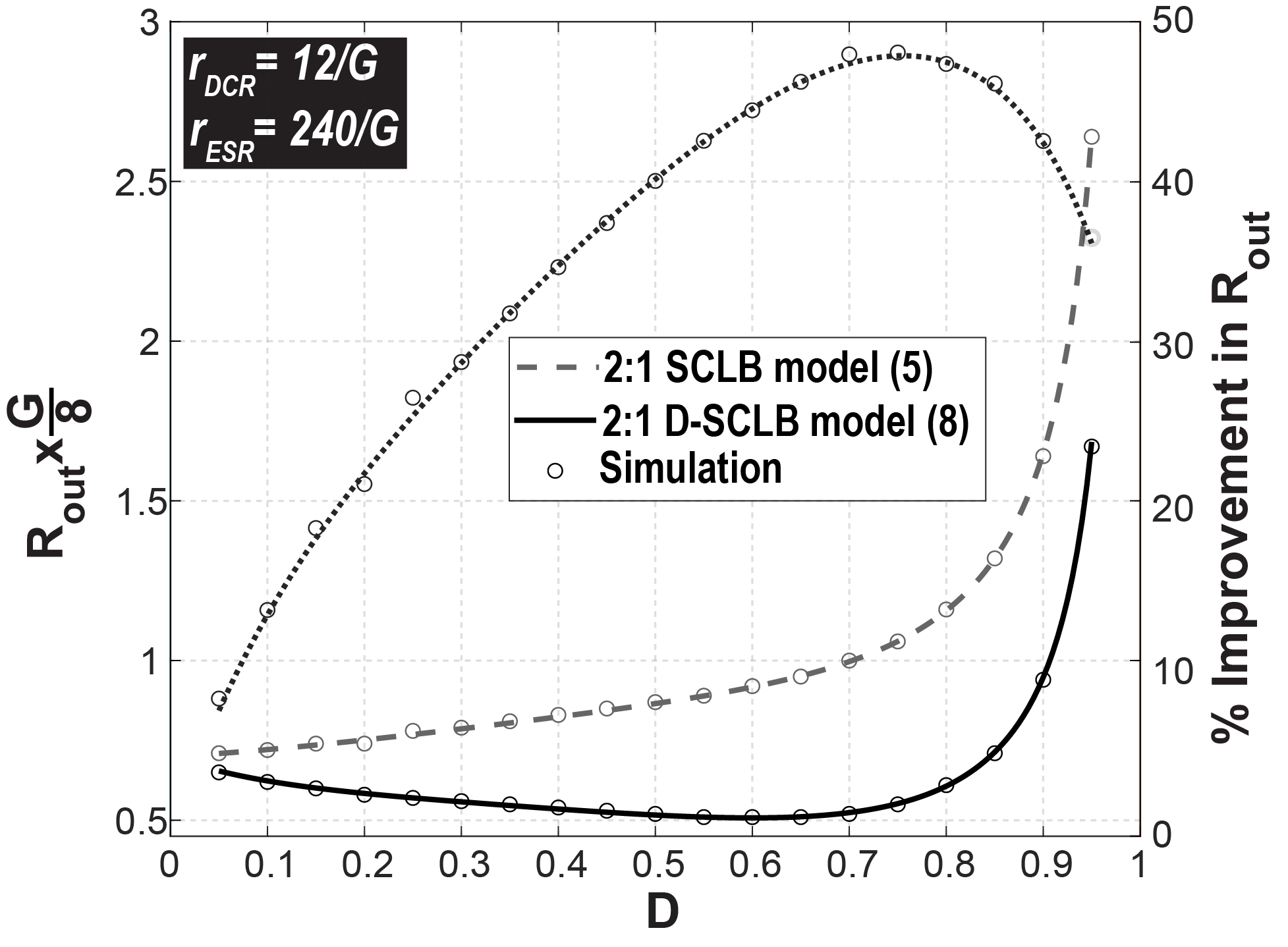}
\caption{Equivalent output impedance of the 2:1 SCLB and 2:1 D-SCLB converters versus duty cycle, $D$, with $r_{DCR}$ = $\frac{12}{G}$ and $r_{ESR}$ = $\frac{240}{G}$. \%Improvement in $R_{out}$ is equal to 100$\times$$(R_{out-DSCLB}-R_{out-SCLB})/R_{out-SCLB}$.}
\label{fig.2x1routcompesr}
\end{figure}

Fig.~\ref{fig.2x1routcompesr} illustrates the equivalent output impedance of the 2:1 SCLB and 2:1 D-SCLB converters as the duty cycle, $D$, is varied from 0 to 1, with $r_{DCR}$\,=\,$\frac{12}{G}$ and $r_{ESR}$\,=\,$\frac{240}{G}$. The $r_{DCR}$ and $r_{ESR}$ values are selected from the datasheet of the inductor used for the measurements~\textcolor{black}{[]}. Because only a fraction of the output current, namely $(1 - M)I_{out}$, flows through the inductor while the remaining current is delivered through the flying capacitor, the normalized output impedance remains less than unity over a wider range of $D$, for instance, up to $D$\,=\,0.9, as opposed to only $D$\,=\,0.7 in Fig.~\ref{fig.2x1routcomp}. Furthermore, the 2:1 D-SCLB converter achieves up to 48\% improvement in $R_{out}$ compared to the 2:1 SCLB, as shown in Fig.~\ref{fig.2x1routcompesr}. The simulation results agree with the model predictions to within 1\%.

\begin{figure}[!tb]
\centering
\includegraphics[width=3.4in]{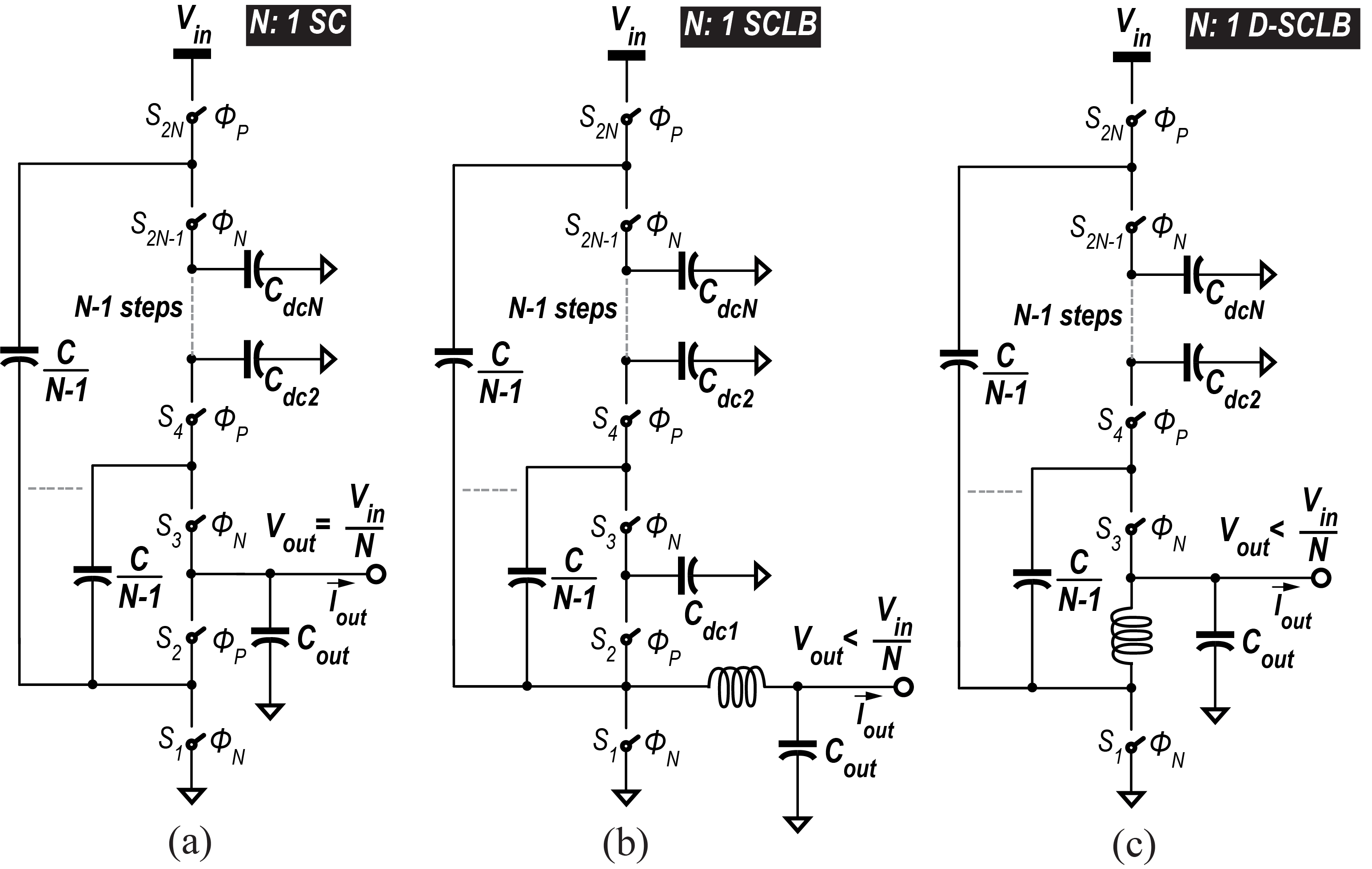}
\caption{Schematic diagrams of the $N$:1 (a) SC, (b) SCLB, and (c) D-SCLB converters.}
\label{fig.Nx1sch}
\end{figure}

\subsection{$N$:1 Converter Output Impedance}
\label{subsec:nx1conoutputimpedance}
The preceding analysis can be extended to $N$:1 SC, SCLB, and D-SCLB converters. An $N$:1 SC converter, shown in Fig.~\ref{fig.Nx1sch}(a), can be derived from the 2:1 SC in Fig.~\ref{fig.2x1scmodel}(b) by dividing the top half-bridge into $N-1$ equal segments and stacking them vertically. Similarly, for optimal conduction losses, an $N$:1 D-SCLB converter, shown in Fig.~\ref{fig.Nx1sch}(c), can be derived from the 2:1 D-SCLB in Fig.~\ref{fig.2x1scmodel}(d) by splitting and vertically stacking the top half-bridge into $N-1$ equal segments. However, this segmentation approach does not yield optimal conduction losses for an $N$:1 SCLB converter.
 \begin{table}[!th]
    \centering
    \caption{Normalized RMS Current ($I_{RMS,i}/I_{out}$) for an $N{:}1$ Converter}
    \renewcommand{\arraystretch}{1.8}
    \setlength{\tabcolsep}{4.5pt}
    \begin{tabular}{c|c|c|c}
        \hline\hline
        \multirow{2}{*}{\textbf{Switch}} & \multicolumn{3}{c}{\textbf{Converter Topology}} \\
        \cline{2-4}
        & \textbf{$N{:}1$ SC} & \textbf{$N{:}1$ SCLB} & \textbf{$N{:}1$ D-SCLB} \\
        \hline
        $S_1$ & $\frac{N-1}{N\sqrt{1-D}}$ &
                $\frac{N-D}{N\sqrt{1-D}}$ &
                $\frac{N-1}{(N-1+D)\sqrt{1-D}}$ \\[6pt]

        $S_2$ & $\frac{N-1}{N\sqrt{D}}$ &
                $\frac{\sqrt{D}}{N}$ &
                -- \\[8pt]

        $S_{3,5,\dots,2N-1}$  &
                $\frac{1}{N\sqrt{1-D}}$ &
                $\frac{D}{N\sqrt{1-D}}$ &
                $\frac{D}{(N-1+D)\sqrt{1-D}}$ \\[8pt]

        $S_{4,6,\dots,2N}$  &
                $\frac{1}{N\sqrt{D}}$ &
                $\frac{\sqrt{D}}{N}$ &
                $\frac{\sqrt{D}}{N-1+D}$ \\[4pt]
        \hline\hline
    \end{tabular}
    \label{tab:rms_currents_N_to_1}
\end{table}

\begin{table}[!th]
    \centering
    \caption{Optimal Switch Relative Sizing for a Fixed Total Conductance, $G$, at $D = 0.5$}
    \renewcommand{\arraystretch}{1.4}
    \setlength{\tabcolsep}{4pt}
    \begin{tabular}{c|c|c|c}
        \hline\hline
        \multirow{2}{*}{\textbf{Switch}}
                 & \multicolumn{3}{c}{\textbf{Converter Topology}} \\ \cline{2-4}
                 & $\boldsymbol{N{:}1}$ \textbf{SC}
                 & $\boldsymbol{N{:}1}$ \textbf{SCLB}
                 & $\boldsymbol{N{:}1}$ \textbf{D-SCLB} \\ \hline
        $S_1$              & $\frac{G}{4}$
                           & $\frac{G}{2}$
                           & $\frac{G}{2}$ \\[6pt]
        $S_2$              & $\frac{G}{4}$
                           & $\frac{G}{2(2N-1)}$
                           & -- \\[6pt]
        $S_{3,5,\dots,2N-1}$
                           & $\frac{G}{4(N-1)}$
                           & $\frac{G}{2(2N-1)}$
                           & $\frac{G}{4(N-1)}$ \\[6pt]
        $S_{4,6,\dots,2N}$
                           & $\frac{G}{4(N-1)}$
                           & $\frac{G}{2(2N-1)}$
                           & $\frac{G}{4(N-1)}$ \\[4pt]
        \hline\hline
    \end{tabular}
    \label{tab:opt_res_Nto1_D05}
\end{table}

The normalized RMS currents ($I_{RMS,i}/I_{out}$) of the switches in the $N$:1 converters are given in Table~\ref{tab:rms_currents_N_to_1}. Using the principle in~\eqref{eq:rsizingfund}, the optimal relative sizes of the switches for the three $N$:1 converters with the same total switch area are summarized in Table~\ref{tab:opt_res_Nto1_D05}. The results in Table~\ref{tab:opt_res_Nto1_D05} confirm the intuitive notion that the top half-bridge of the 2:1 SC and 2:1 D-SCLB converters can be split into $N-1$ equal segments and stacked vertically to form their respective $N$:1 converters; however, this splitting approach does not yield optimal conduction losses for the SCLB converter.

The equivalent output impedance of the $N$:1 SC converter is given by
{%
\setlength{\abovedisplayskip}{0.5\baselineskip}%
\setlength{\belowdisplayskip}{0.5\baselineskip}%
\begin{equation}
R_{out} = \frac{1}{G}\times\frac{8(N-1)^2}{N^2}\times\frac{1}{D(1-D)}.
\label{eq:routNx1sc}
\end{equation}}

The ideal conversion ratio of an $N$:1 SCLB converter, shown in Fig.~\ref{fig.Nx1sch}(b), is given by
{%
\setlength{\abovedisplayskip}{0.5\baselineskip}%
\setlength{\belowdisplayskip}{0.5\baselineskip}%
\begin{equation}
M = \frac{D}{N}.
\end{equation}}

The inductor average and ripple currents can be expressed as
{%
\setlength{\abovedisplayskip}{0.5\baselineskip}%
\setlength{\belowdisplayskip}{0.5\baselineskip}%
\begin{equation}
\begin{aligned}
I_L &= I_{out}, \\
\Delta I &= \frac{D(1-D)}{N}\frac{V_{in}}{fL}.
\end{aligned}
\end{equation}}

Thus, the equivalent output impedance of an $N$:1 SCLB converter is given by
{%
\setlength{\abovedisplayskip}{0.5\baselineskip}%
\setlength{\belowdisplayskip}{0.5\baselineskip}%
\begin{equation}
\begin{aligned}
R_{out} &= r_{DCR}\\
&+ \frac{2}{G}\left(\frac{N^{2}+N(2N-3)D-2(N-1)D^{2}}{N^{2}(1-D)}\right)\\
&+ \frac{\Delta I^2}{12\,I_{out}^{2}}\left(r_{ESR}+\frac{2}{G}\left(\frac{N+(N-1)D}{N}\right)\right).
\end{aligned}
\label{eq:routNx1sclb}
\end{equation}}

The ideal conversion ratio of an $N$:1 D-SCLB converter, shown in Fig.~\ref{fig.Nx1sch}(c), is given by
{%
\setlength{\abovedisplayskip}{0.5\baselineskip}%
\setlength{\belowdisplayskip}{0.5\baselineskip}%
\begin{equation}
M = \frac{D}{N-1+D}.
\end{equation}}

The corresponding average and ripple currents in the inductor are given by
{%
\setlength{\abovedisplayskip}{0.5\baselineskip}%
\setlength{\belowdisplayskip}{0.5\baselineskip}%
\begin{equation}
\begin{aligned}
I_L &= (1-M)I_{out}, \\
\Delta I &= \frac{D(1-D)}{N-1+D}\frac{V_{in}}{fL}.
\end{aligned}
\end{equation}}

The expression for the equivalent output impedance given by~\eqref{eq:rout2x1dsclb} is also valid for an $N$:1 D-SCLB converter.

\subsection{Performance Comparison for a 48\,V to Point-of-Load Voltage Conversion}

\begin{figure}[!tb]
\centering
\includegraphics[width=2.4in]{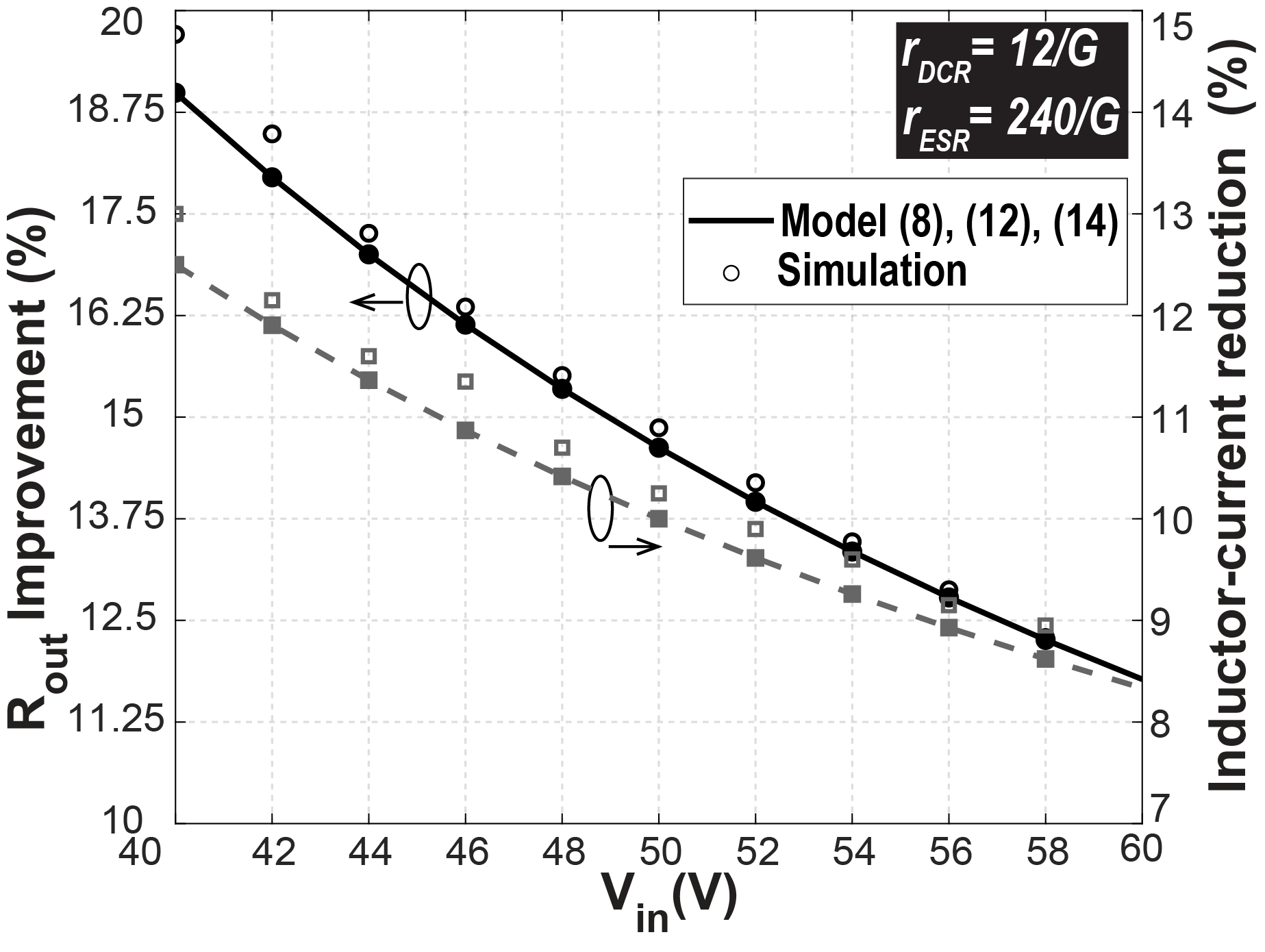}
\caption{Percentage improvement in equivalent output impedance and reduction in average inductor current of the 4:1 D-SCLB converter versus the 4:1 SCLB converter, for a constant 9\,V output with $V_\mathrm{in}$ from 40\,V to 60\,V.}
\label{fig.4x1routimp}
\end{figure}

The output-impedance analysis in Section~\ref{subsec:nx1conoutputimpedance} is utilized to evaluate the performance advantages of the proposed 4:1 D-SCLB converter relative to a 4:1 SCLB converter for 48\,V-to-point-of-load voltage conversion suitable for data-center applications. With optimal relative switch sizing, $r_{DCR}=3\times r_{ON}$ and $r_{ESR}=30\times r_{ON}$, the percentage improvement in the equivalent output impedance and the reduction in the average inductor current of the 4:1 D-SCLB converter, while generating a constant 9\,V output from an input voltage varying between 40\,V and 60\,V, are illustrated in Fig.~\ref{fig.4x1routimp}. The proposed 4:1 D-SCLB converter achieves up to a 20\% reduction in the average inductor current and up to a 30\% improvement in the equivalent output impedance compared to the 4:1 SCLB converter.
\begin{figure}[!tb]
\centering
\includegraphics[width=2.4in]{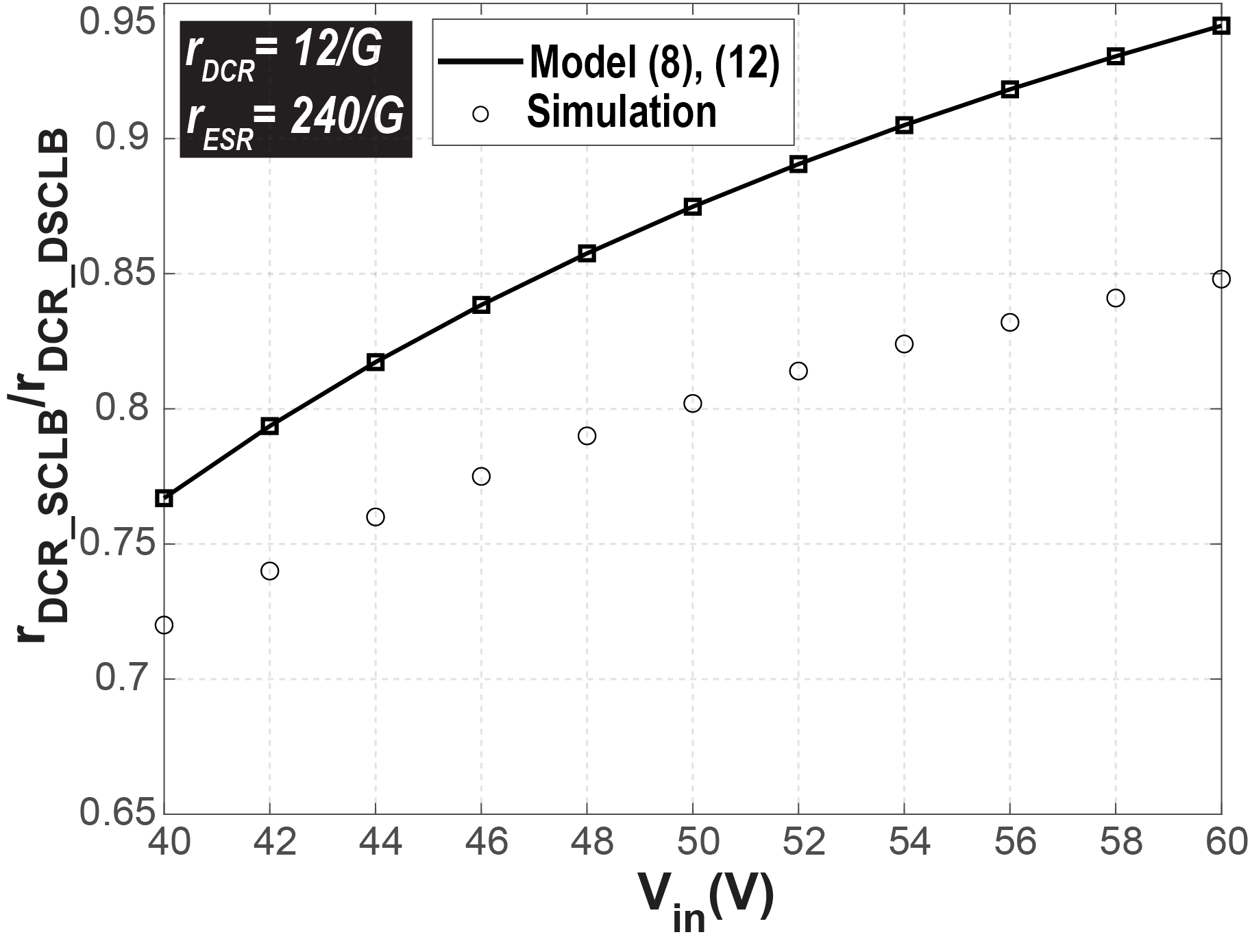}
\caption{Ratio of inductor DCR between the 4:1 SCLB and 4:1 D-SCLB converters for equal output impedance.}
\label{fig.4x1ksroutmatch}
\end{figure}
To achieve the same converter performance between the 4:1 SCLB and 4:1 D-SCLB converters, a larger inductor with lower $r_{DCR}$ and $r_{ESR}$ values needs to be chosen with the SCLB converter. For example, with a design employing $r_{ON}=24\,m\Omega$, $V_{in}=48\,V$, $f$\,=\,500\,kHz, $L$=\,10\,$\mu$H, and $R_{load}$\,=\,9\,$\Omega$, the SCLB converter $r_{DCR}$ and $r_{ESR}$ should be 0.57$\times$ those of the D-SCLB converter to achieve the same conversion efficiency, as illustrated in Fig.~\ref{fig.4x1ksroutmatch}.

\subsection{$N$:1 Converter Output Impedance and Performance Comparison for a 48\,V to Point-of-Load Voltage Conversion with Uniform Switch Sizing}
In this section, the output impedances of the $N$:1 SCLB and $N$:1 D-SCLB converters are derived, and a performance comparison is presented for the case in which uniform switch sizing is chosen for design simplicity or because switches with the required on-resistance values are unavailable. 
In this case, the output impedance of an $N$:1 SCLB is given by
{%
\setlength{\abovedisplayskip}{0.5\baselineskip}%
\setlength{\belowdisplayskip}{0.5\baselineskip}%
\begin{equation}
\begin{aligned}
R_{out}&=r_{DCR}\\
&+\frac{2}{G}\left(D
+\frac{(N-1)D^{2}}{N(1-D)}
+\frac{(N-D)^{2}}{N(1-D)}\right)\\
&+\frac{\Delta I^{2}}{12\,I_{out}^{2}}
\left(r_{ESR}+\frac{2}{G}\left(N(1-D)+D\right)\right).
\label{eq:routNx1sclb_uss}
\end{aligned}
\end{equation}}
The output impedance of an $N$:1 D-SCLB is given by
{%
\setlength{\abovedisplayskip}{0.5\baselineskip}%
\setlength{\belowdisplayskip}{0.5\baselineskip}%
\begin{equation}
\begin{aligned}
R_{out} &= (1-M)^2\left(\;r_{DCR} + \frac{1}{G}\left(
      \frac{(2N-1)(D+N-1)}{(N-1)(1-D)}\right)\right) \\
&+\frac{\Delta I^{2}}{12\,I_{out}^{2}}\left(r_{ESR}
        +\frac{1}{G}\Bigl(\frac{(2N-1))((N-1)(1-D)+D)}{N-1}\Bigr)\right).
\end{aligned}
\end{equation}}

\begin{figure}[!tb]
\centering
\includegraphics[width=2.4in]{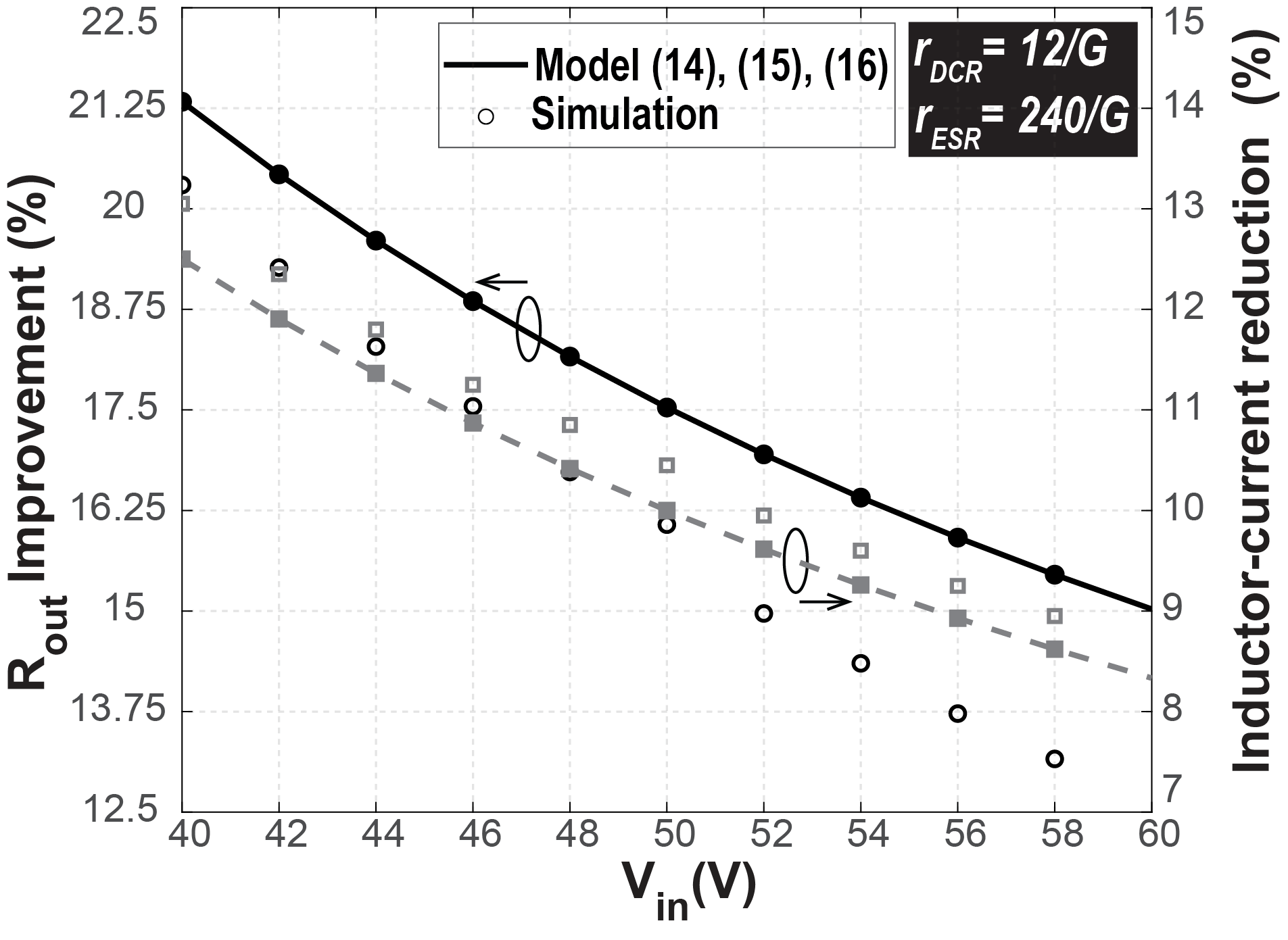}
\caption{With uniform switch sizing, the percentage improvement in equivalent output impedance and reduction in average inductor current of the 4:1 D-SCLB converter versus the 4:1 SCLB converter, for a constant 9\,V output with $V_\mathrm{in}$ from 40\,V to 60\,V.}
\label{fig.4x1routimp_uss}
\end{figure}

\begin{figure}[!tb]
\centering
\includegraphics[width=2.4in]{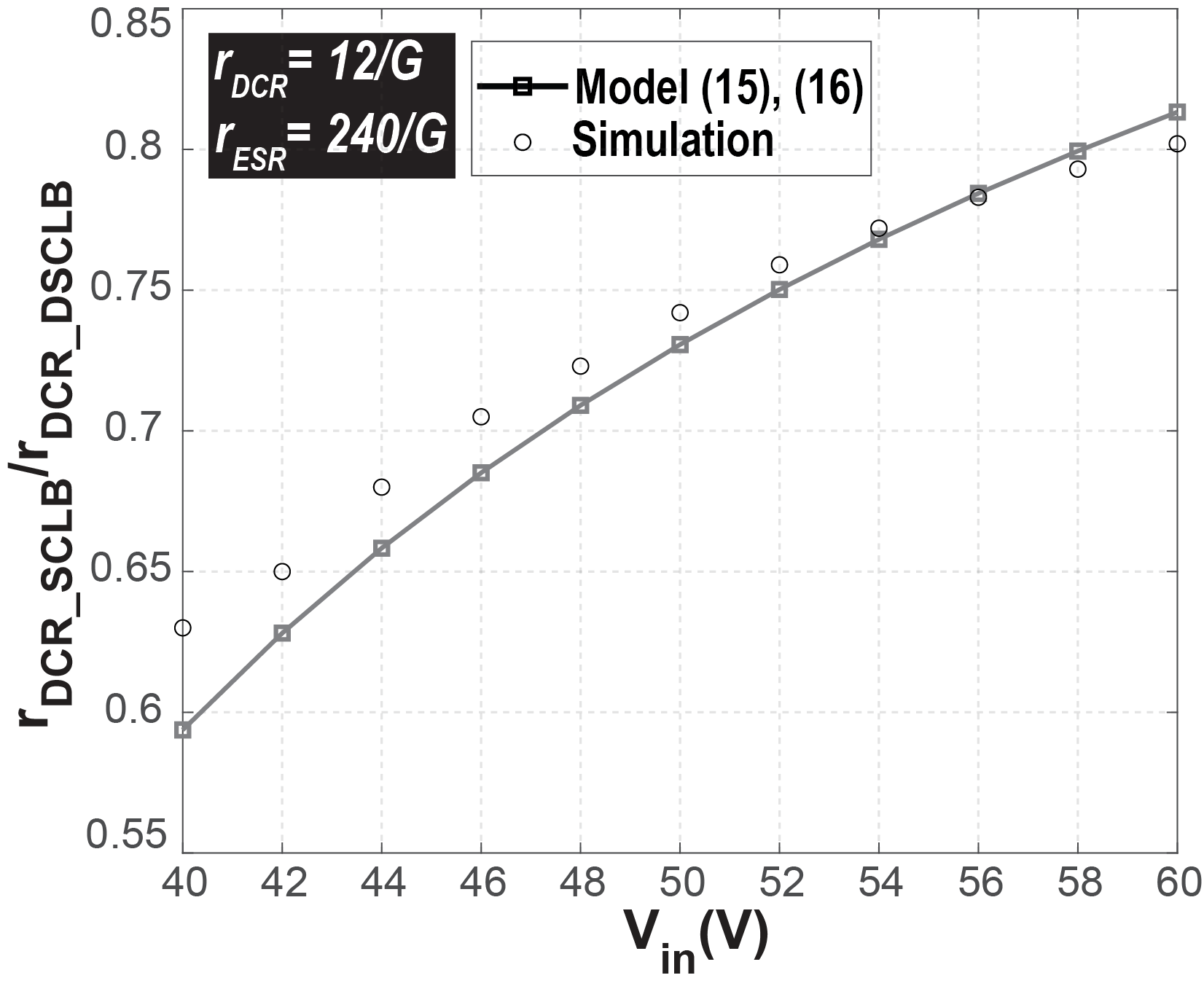}
\caption{With uniform switch sizing, the ratio of inductor DCR between the 4:1 SCLB and 4:1 D-SCLB converters for equal output impedance.}
\label{fig.4x1ksroutmatchuss}
\end{figure}
Under uniform switch sizing with $r_{DCR} = 3 \times r_{ON}$ and $r_{ESR} = 30 \times r_{ON}$, the percentage improvement in equivalent output impedance and the reduction in average inductor current for the 4:1 D-SCLB converter—while delivering a constant 9\,V output from a 40–60\,V input, are shown in Fig.~\ref{fig.4x1routimp_uss}. The proposed 4:1 D-SCLB achieves up to a 8\% improvement in equivalent output impedance and up to a 6\% reduction in average inductor current compared to the 4:1 SCLB converter. To achieve comparable performance between the 4:1 SCLB and 4:1 D-SCLB converters under uniform switch sizing, a larger inductor with 4$\times$ lower $r_{DCR}$ and $r_{ESR}$ values must be used in the SCLB converter, as illustrated in Fig.~\ref{fig.4x1ksroutmatchuss}.

\section{Hardware Design and Implementation}

\begin{figure}[!tb]
\centering
\includegraphics[width=3.5in]{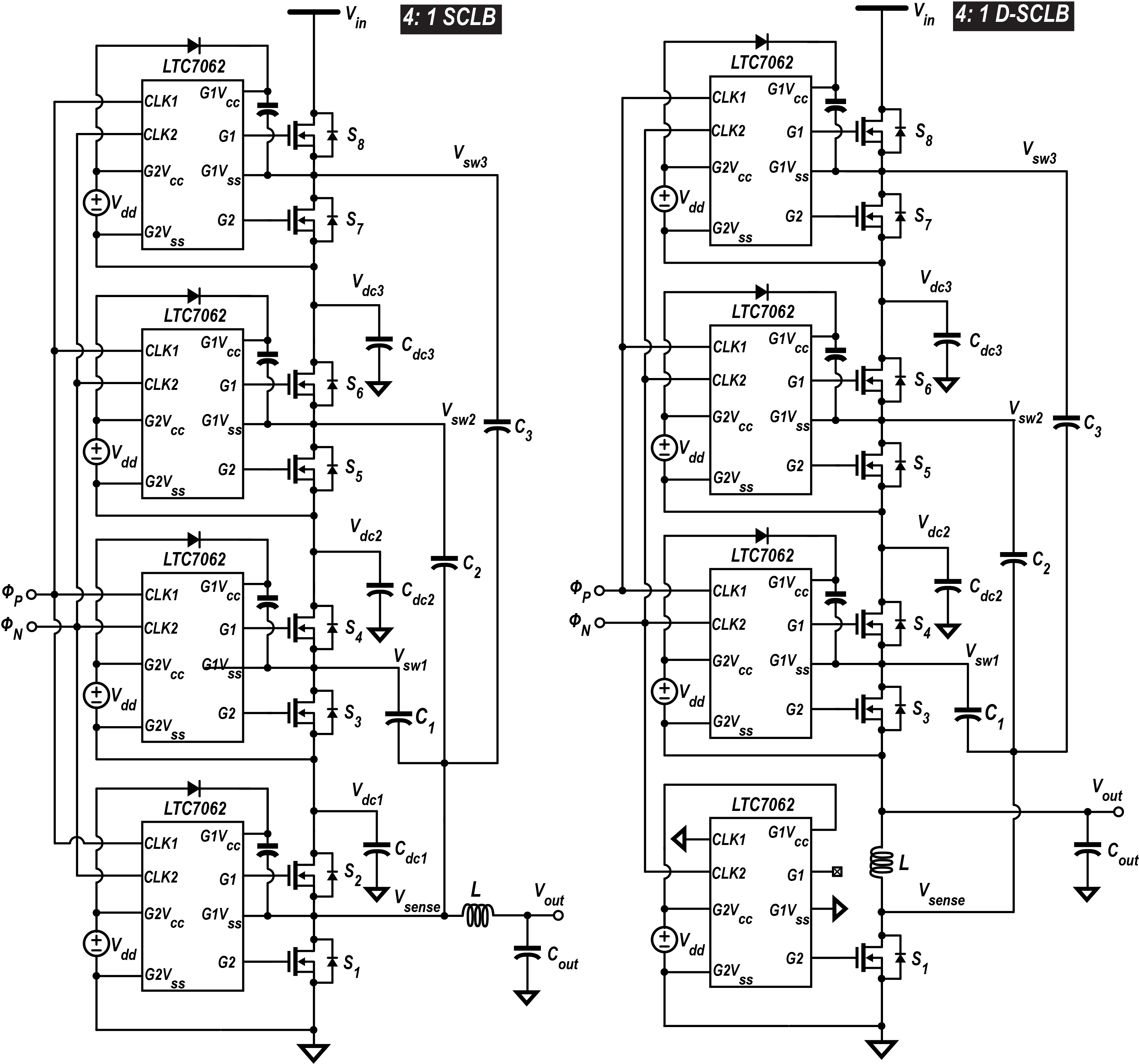}
\caption{Schematic of the implemented 4:1 SCLB and D-SCLB converters.}\label{fig.Nx1DSCLB_schematic}
\end{figure}

\begin{figure}[!tb]
\centering
\includegraphics[width=3.5in]{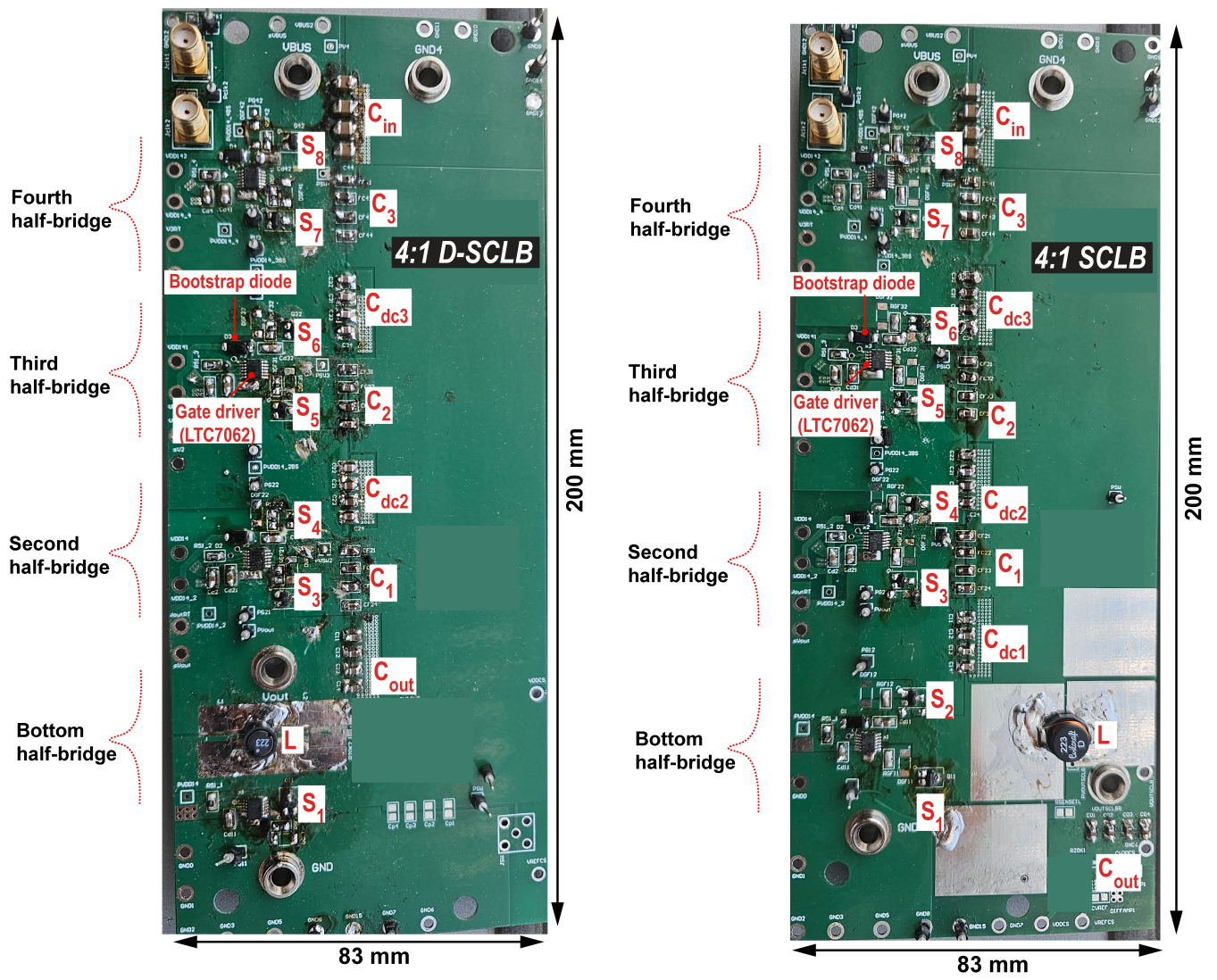}
\caption{Printed circuit board design.}\label{fig.PCB}
\end{figure}

In order to experimentally validate the advantages of the proposed $N$:1 D-SCLB converter over an $N$:1 SCLB converter, 4:1 converter prototypes are implemented on a standard two-layer (2 oz copper), 0.6\,mm thick, low-cost PCB. The prototypes are designed for an input voltage range of 40–60\,V and a fixed output voltage of 8\,V. The schematics of the implemented 4:1 SCLB and D-SCLB converters are shown in Fig.\ref{fig.Nx1DSCLB_schematic}(a) and (b), respectively, while Fig.\ref{fig.PCB}(a) and (b) show the corresponding prototype PCBs, including connectors and additional filter capacitors. The key components are highlighted in the figure, and their specifications are summarized in Table~\ref{tab:component_list}. A 22\,$\mu$H inductor with a DCR of 320\,m$\Omega$ is used to demonstrate the performance improvement of the proposed converter over the SCLB converter when both converters employ the same inductor. Additionally, a 22\,$\mu$H inductor with a lower DCR of 167\,m$\Omega$ is used in the SCLB converter to achieve a conversion efficiency comparable to that of the proposed converter.
\begin{table*}[!th]
    \centering
    \caption{Component List of the Hardware Prototype}
       \renewcommand{\arraystretch}{1.5}
    \setlength{\tabcolsep}{10pt}
    \begin{tabular}{c|c|c|c}
        \hline
        \hline
        \textbf{Component} & \textbf{Converter}& \textbf{Part number} & \textbf{Parameters} \\
        \hline
         MOSFET $S_{1,3-8}$ &SCLB and D-SCLB & Infineon IRFML8244TRPbF & 25 V, 24 m$\Omega$ \\
        \hline
         MOSFET $S_{2}$ &SCLB& Infineon IRFML8244TRPbF & 25 V, 24 m$\Omega$ \\
        \hline
         Flying capacitor $C_{1-3}$ &SCLB and D-SCLB & Murata GRM21BR61H106KE43K & X5R, 50 V, 10 \textmu F$\times$4 (in parallel) \\
         Inductor $L$ &D-SCLB & Coilcraft MSS7348-223 & 22 $\mu$H, 74 m$\Omega$, 2.3 A \\
         Inductor $L$ &SCLB& Coilcraft CSHB0540R-220M & 22 $\mu$H, 64 m$\Omega$, 2.7 A \\
        \hline
         Input capacitor $C_{in}$ &SCLB and D-SCLB & Murata GRM32EC72A106KE05K & X7S, 100 V, 10 \textmu F$\times$4 (in parallel)\\
         Output capacitor $C_{out}$ &SCLB and D-SCLB & Murata GRM21BR61E106KA73L & X5R, 25 V, 10 \textmu F$\times$4 (in parallel) \\
         Static ladder capacitor $C_{dc1}$ &SCLB & Murata GRM21BR61E106KA73L & X5R, 25 V, 10 \textmu F$\times$4 (in parallel) \\
         Static ladder capacitor $C_{dc2-dc3}$ &SCLB and D-SCLB & Murata GRM21BR61H106KE43K & X5R, 50 V, 10 \textmu F$\times$4 (in parallel) \\
         \hline
        Gate driver &SCLB and D-SCLB & Analog Devices LTC7062 & Dual high-side gate driver, 100V \\
        Bootstrap diode &SCLB and D-SCLB & Rohm RSX205LAM30TR & Schottky diode, 30V\\
        \hline
        \hline
    \end{tabular}
    
    \label{tab:component_list}
\end{table*}

Class II ceramic capacitors (X5R and X7S) offer high energy density comparable to that of electrolytic capacitors~\cite{b13},\cite{b14}. Accordingly, they are employed as flying ladder capacitors ($C_{1-3}$) and as decoupling capacitors ($C_{in}$, $C_{out}$, and $C_{dc2-3}$) for both the converters shown in Fig.\ref{fig.Nx1DSCLB_schematic}. Additionally, the SCLB converter require a $C_{dc1}$ decoupling capacitor, as shown in Fig.\ref{fig.Nx1DSCLB_schematic}(a). In a ladder converter, the DC voltage at each node increases in integer multiples of $(V_{in} - V_{out}) / (N - 1)$ relative to the output voltage. Capacitor voltage ratings are selected based on availability and the required voltage margin. At the output node, a 25\,V capacitor (approximately 3$\times$ the operating voltage, 8\,V) is used for $C_{out}$. For the intermediate ladder nodes, 50\,V and 25\,V capacitors (approximately 2$\times$ and 1.33$\times$ the voltage ratings) are used for $C_{dc2}$ and $C_{dc3}$, respectively. At the input node, a 100\,V capacitor (approximately 2$\times$ the nominal input voltage, 48\,V) is selected for $C_{in}$. Similarly, for the flying capacitors, voltage ratings of approximately 3$\times$, 2$\times$, and 1.33$\times$ are chosen for $C_1$, $C_2$, and $C_3$, respectively. For the additional $C_{dc1}$ required for the SCLB converter, a 25\,V capacitor (approximately 2$\times$ the nominal operating voltage, 12\,V) is used. Due to the voltage-dependent capacitance reduction inherent to Class II dielectrics, the effective capacitance decreases with increasing DC bias. To ensure sufficient decoupling (suppressing voltage ripple to negligible levels) and to avoid operating in the slow-switching limit at the selected switching frequency, an array of four capacitors is connected in parallel at each node.%

For switch selection, the inherent voltage division property of the SC ladder topology ensures that each switch only needs to block a fraction of the input voltage, \mbox{$(V_{in} - V_{out})/(N - 1)$}, enabling the use of low-voltage MOSFETs rated at 25\,V. Accordingly, seven identical MOSFETs, $S_{1,3\text{--}8}$, are employed in the D-SCLB converter, each with an on-resistance of $r_{ON}$\,=\,24\,m$\Omega$ (specified at $V_{GS}$\,=\,6\,V and $I_D$\,=\,5.8\,A), as shown in Fig.\ref{fig.Nx1DSCLB_schematic}(b). An additional switch, $S_2$, using the same device, is included in the SCLB converter, as shown in Fig.\ref{fig.Nx1DSCLB_schematic}(a).%

Two low-voltage, ground-referenced, out-of-phase external clocks, $\phi_P$ and $\phi_N$, as shown in Fig.\ref{fig.Nx1DSCLB_schematic}, are generated with controlled dead time using a Tektronix AFG31052 two-channel synchronous signal generator. An LTC7062 IC\cite{LTC7062}, featuring a symmetric floating gate driver architecture with an integrated level shifter, is used to drive each of the $N$ half-bridge inverter switch pairs. This IC provides symmetric gate drive signals for both the high-side and low-side MOSFETs in each half bridge. The gate driver for the bottom MOSFET in each stacked half bridge is powered by a 6\,V supply referenced to the local ground of the half bridge, defined by the static ladder node voltages $V_{out}$, $V_{dc1}$, $V_{dc2}$, and $V_{dc3}$. In contrast, the top MOSFET gate driver in each half bridge is powered by an approximately 6\,V bootstrapped supply referenced to the corresponding switching node, namely $V_{sense}$, $V_{sw1}$, $V_{sw2}$, and $V_{sw3}$. In the D-SCLB converter, the bottom gate driver drives the ground-referenced switch $S_1$, while the top-side driver remains unused. While the board layout is optimized to minimize parasitic effects, component placement is not optimized for compactness, as the primary objective of the prototype is to experimentally validate the efficacy of the proposed topology, which inherently achieves reduced average inductor current. The design is scalable to higher conversion ratios by extending the ladder with additional stages. This scalability is facilitated by the PCB layout, which supports such extension with minimal increase in routing or structural complexity, as illustrated in Fig.~\ref{}.

The volumes of the passive components used in the D-SCLB and SCLB converters to achieve similar conversion efficiency are compared in Table~\ref{tab:passive_volume}. The reduced conduction losses associated with inductor DCR in the D-SCLB topology enable the use of a smaller-volume inductor with higher DCR, compared to the inductor required in the SCLB converter under identical operating conditions. As shown in Table~\ref{tab:passive_volume}, the inductor volume ratio (D-SCLB to SCLB) is 0.49, and the ratio of the total passive (inductor and capacitors) component volume is 0.505.
\begin{table}[h]
    \centering
    \caption{Comparison of Passive Component Volumes for D-SCLB and SCLB Converters at Similar Conversion Efficiency}
    \renewcommand{\arraystretch}{1.2}
    \setlength{\tabcolsep}{6pt}
    \begin{tabular}{c|c|c|c}
        \hline
        \hline
        \textbf{} & \textbf{SCLB} & \textbf{D-SCLB}& \textbf{Ratio (D-SCLB to SCLB)}\\
        \hline
         Inductor (mm\textsuperscript{3}) &  256  & 831 & 3.25 \\
         Capacitor (mm\textsuperscript{3}) & 43 & 43 & 1 \\
         Total (mm\textsuperscript{3})& 299  & 874 & 2.92 \\
        \hline
        \hline
    \end{tabular}
    \vspace{0.5em}
    \begin{minipage}{1\linewidth}
\justifying
\noindent

\end{minipage}
    \label{tab:passive_volume}
\end{table}

An additional advantage of using an inductor with relatively high DCR is the reduction in inductor height, which helps lower the overall profile of the prototype. This reduction has a significant impact on the total volume, as the switches and control circuitry typically occupy minimal vertical space. Furthermore, the D-SCLB converter requires one fewer switch ($S_1$) and decoupling capacitor ($C_{dc1}$) compared to the SCLB converter. As a result, the overall prototype volume of the D-SCLB is reduced, with a D-SCLB to SCLB volume ratio of 0.481, as detailed in Table~\ref{tab:total_volume}.
\begin{table}[h]
    \centering
    \caption{Overall Volume of the Prototypes}
    \renewcommand{\arraystretch}{1.2}
    \setlength{\tabcolsep}{3.3pt}
    \begin{tabular}{c|c|c|c}
        \hline
        \hline
        \textbf{} & \textbf{D-SCLB} & \textbf{SCLB}& \textbf{Ratio (SCLB to D-SCLB)}\\
        \hline
         Passive components (mm\textsuperscript{3})  &   458 & 1019 & 2.22\\
         Switches (mm\textsuperscript{3})&  48 & 55 & 1.15\\
         Gate driver  (mm\textsuperscript{3})& 79  & 79 & 1\\
         Bootstrap diode (mm\textsuperscript{3}) &  33 & 44 & 1.36 \\
         Total volume  (mm\textsuperscript{3})&  618 & 1198 & 1.94\\
        \hline
        \hline
    \end{tabular}
    \vspace{0.5em}
    \begin{minipage}{1\linewidth}
\justifying
\noindent

\end{minipage}
    \label{tab:total_volume}
\end{table}

\begin{table}[h]
    \centering
    \caption{Key Parameters of the Hardware Prototypes}
    \renewcommand{\arraystretch}{1.2}
    \setlength{\tabcolsep}{24pt}
    \begin{tabular}{c|c}
        \hline
        \hline
        \textbf{Parameter} & \textbf{Value} \\
        \hline
         Input voltage & 40 - 60\,V \\
         Tested output voltage & 5\,V \\
         Load resistance & 2.5 - 25\,$\Omega$ \\
         Switching frequency &  250\,kHz\\
         Gate drive voltage & 6\,V \\
        \hline
        \hline
    \end{tabular}
    \vspace{0.5em}
    \label{tab:key_parameter}
\end{table}

\section{Experimental Verification}
Table~\ref{tab:key_parameter} summarizes the key specifications of the measured converter. The prototype was evaluated over a load range of 2.5$\Omega$ to 25$\Omega$. A Keysight EL34143A DC electronic load, configured in four-wire sense mode, was used to precisely set the resistive load. The input current was measured using a B\&K Precision 5493C 6.5-digit digital multimeter to ensure accurate efficiency calculation.
\begin{figure}[!tb]
\centering
\includegraphics[width=2.4in]{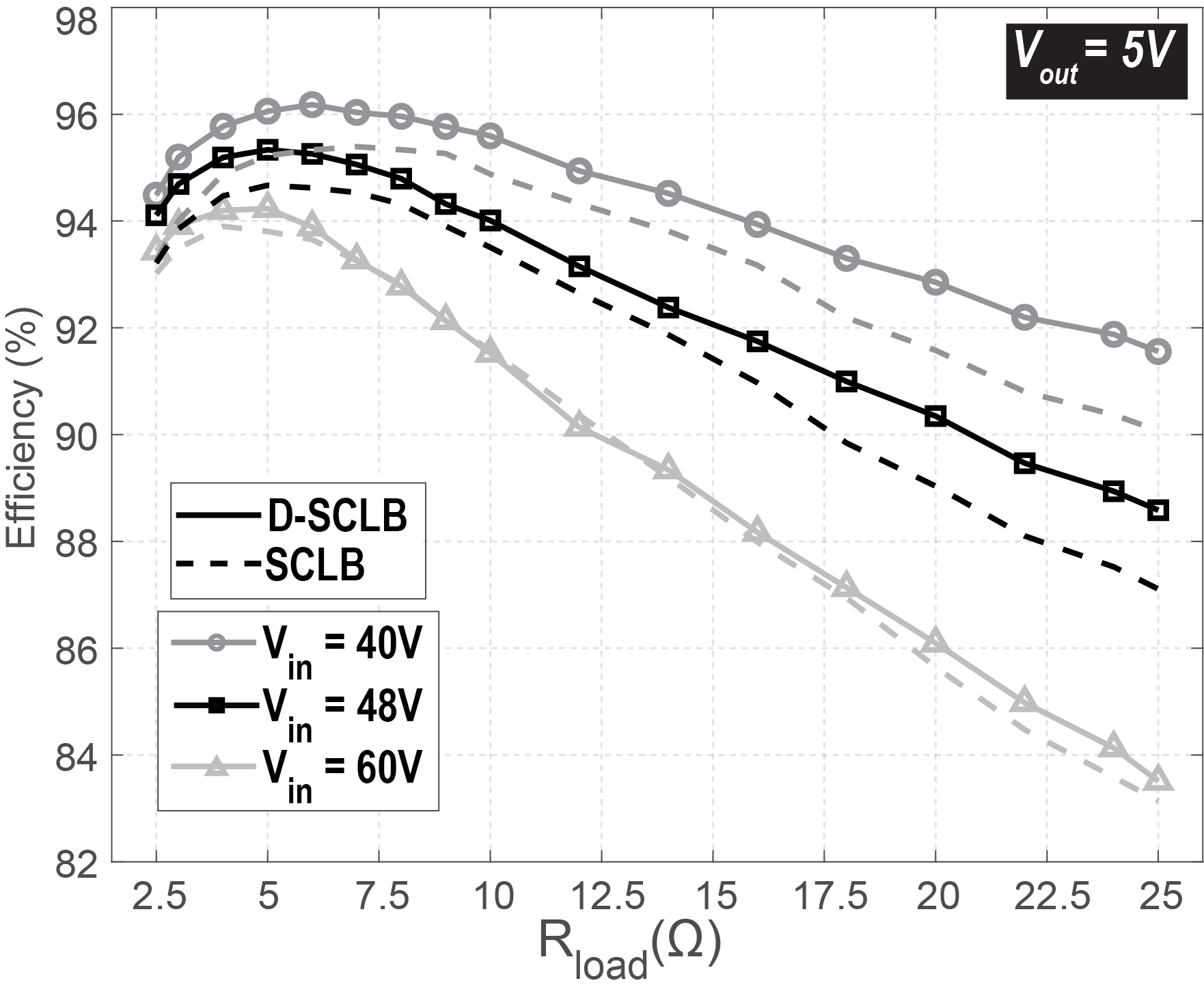}
\caption{Efficiency profile comparison.}
\label{fig.eff_prof_comp}
\end{figure}


\begin{figure}[!tb]
    \centering
    \subfloat[]{%
        \includegraphics[width=2.4in]{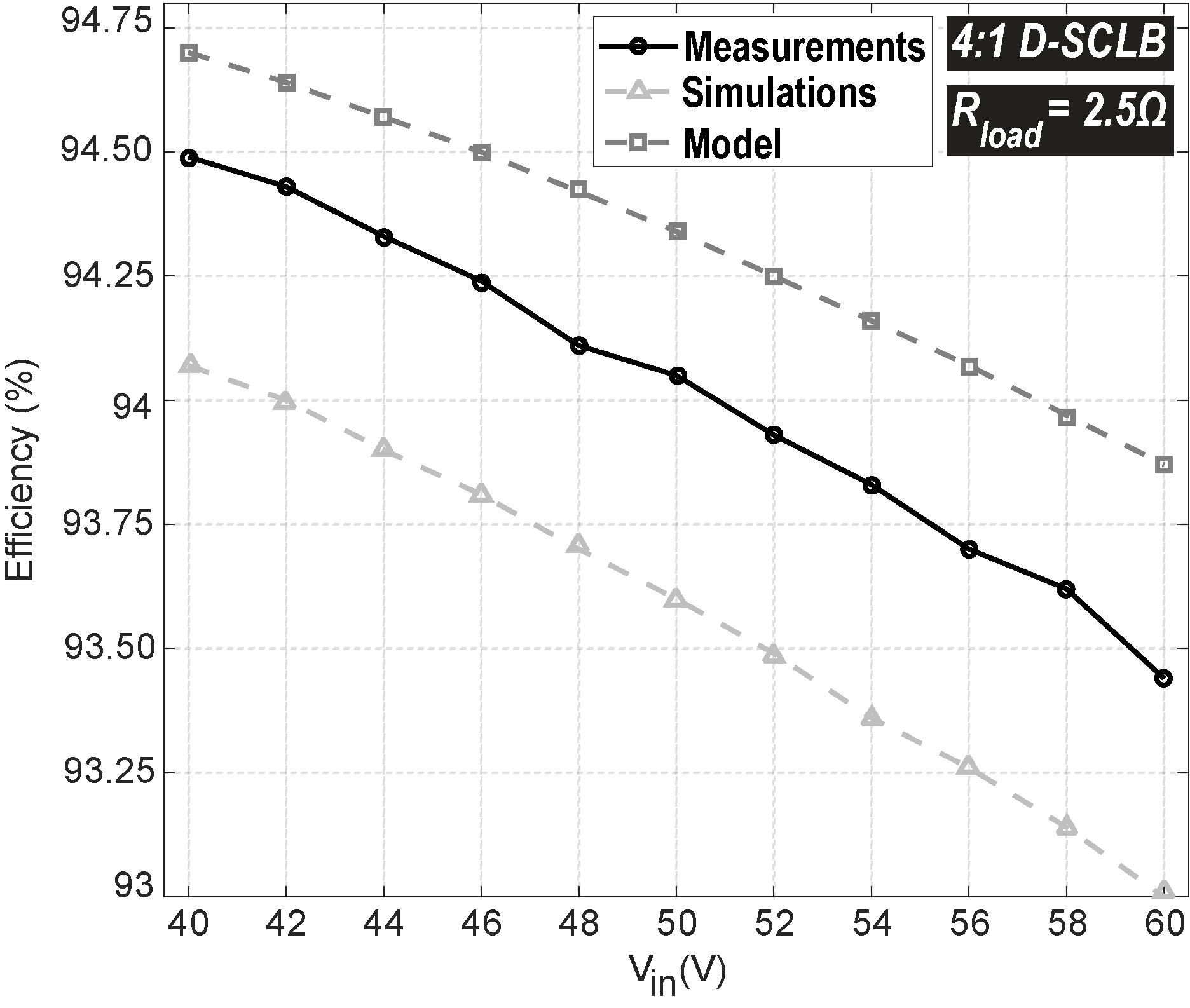}
        \label{fig.eff_prof_comp_dsclb_2p5ohms}
    }\\
    \subfloat[]{%
        \includegraphics[width=2.39in]{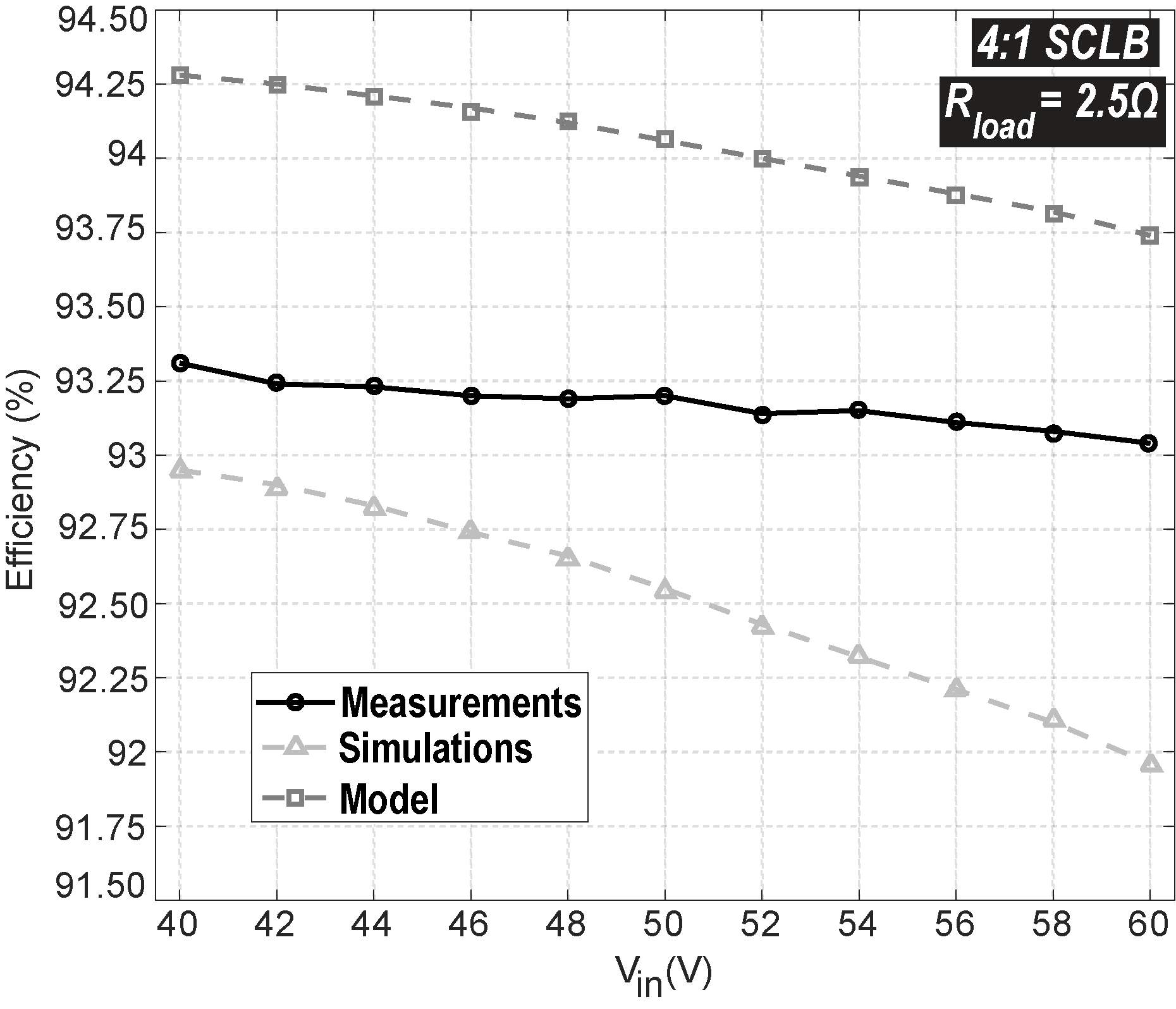}
        \label{fig.eff_prof_comp_sclb_2p5ohms}
    }\\
     \subfloat[]{%
        \includegraphics[width=2.3in]{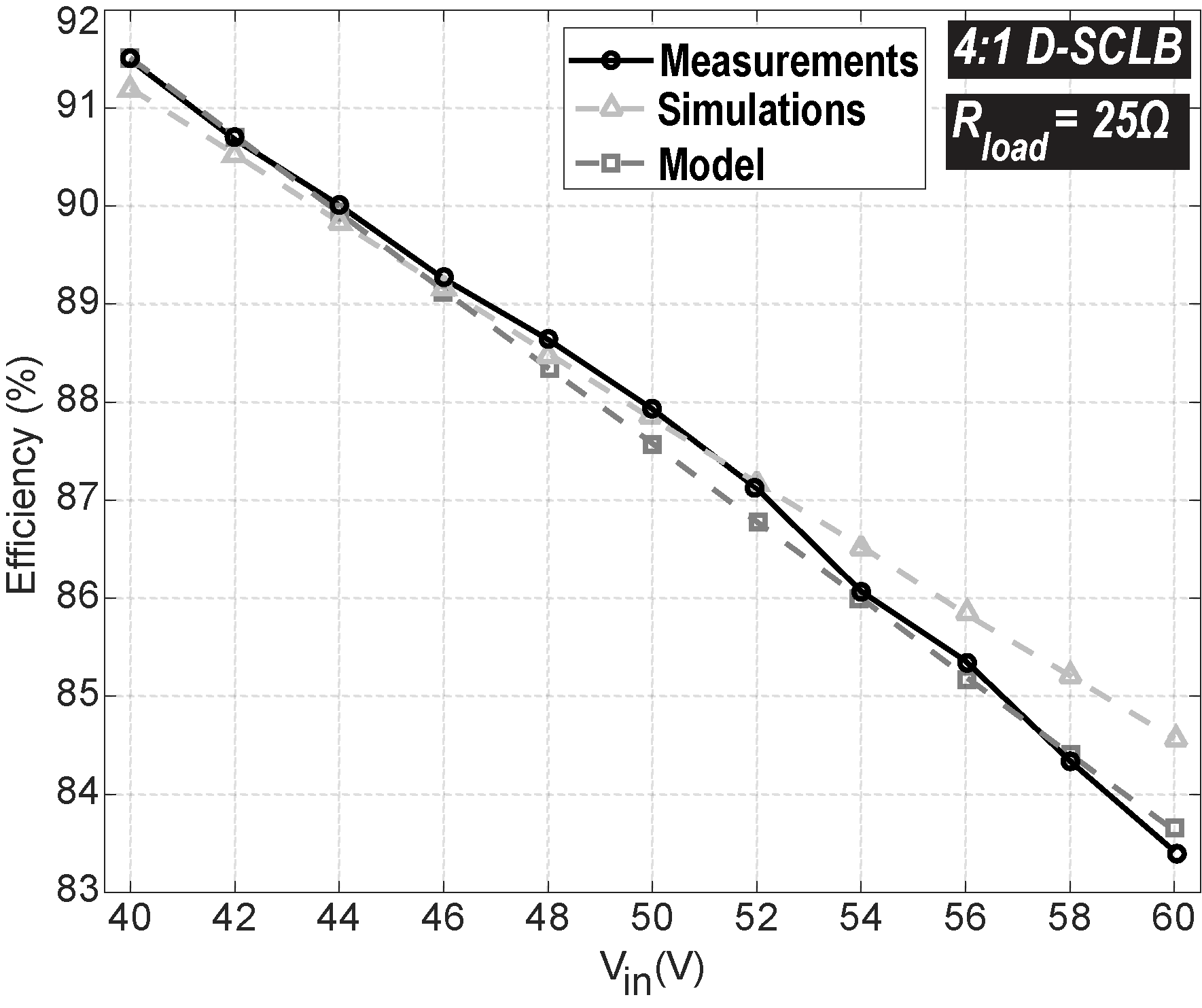}
        \label{fig.eff_prof_comp_dsclb_2p5ohms}
    }\\
    \subfloat[]{%
        \includegraphics[width=2.35in]{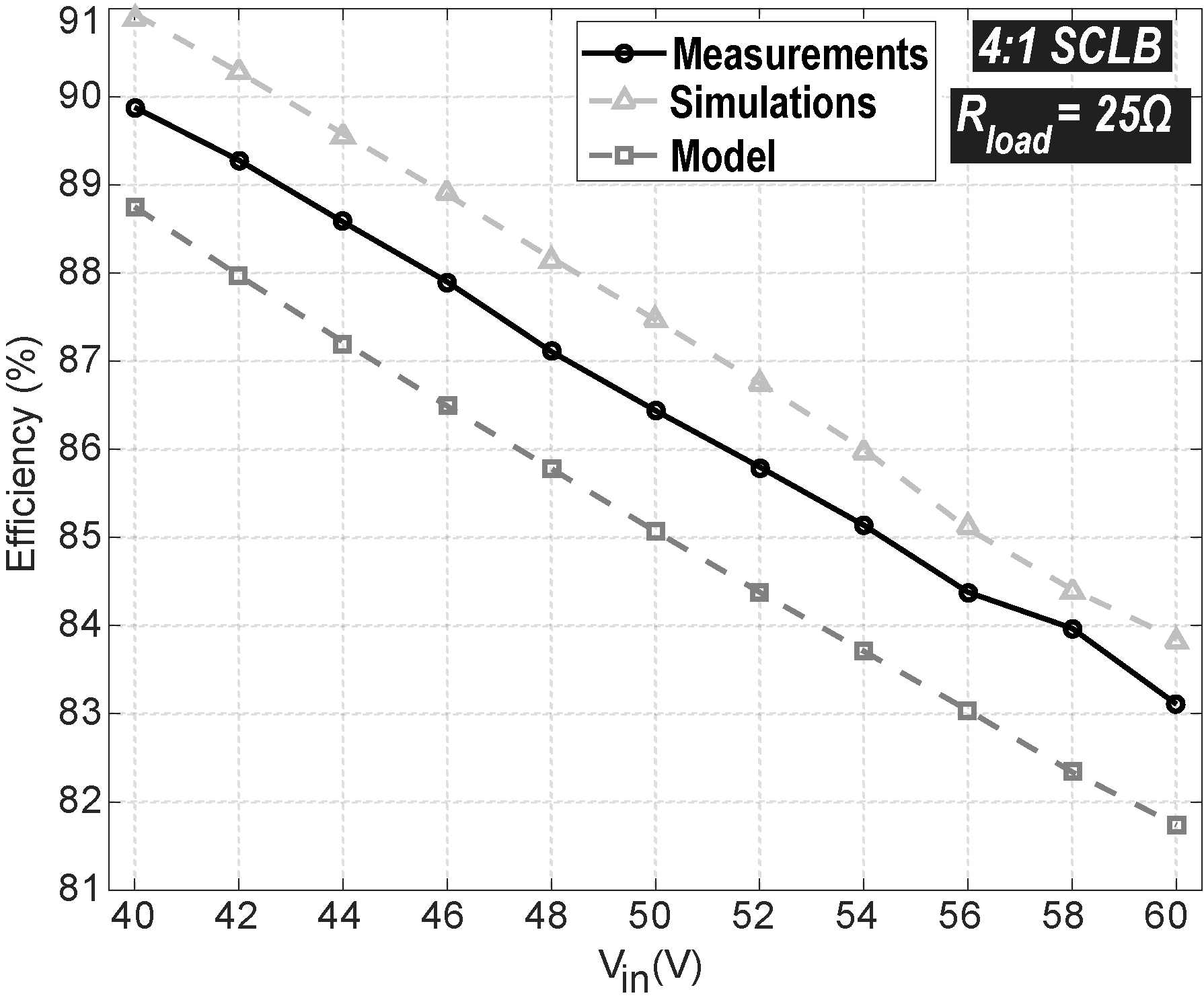}
        \label{fig.eff_prof_comp_sclb_2p5ohms}
    }
    \caption{Efficiency profile comparison between DSCLB and SCLB converters.}
    \label{fig.eff_prof_comp_vin}
\end{figure}

The measurement results in Fig.~\ref{fig.eff_prof_comp} illustrate the 4:1 D-SCLB and SCLB converters’ efficiency as a function of load resistance for a fixed 5\,V output and input voltages of 40\,V, 48\,V, and 60\,V. The D-SCLB converter with a 74\,m$\Omega$ DCR inductor achieves a peak efficiency of 96.2\% and maintains above 83.5\% efficiency across the entire range of input voltages and load resistances. The SCLB converter, employing a 64\,m$\Omega$ DCR inductor, reaches a peak efficiency of 95.4\% and sustains above 83.1\% under the same conditions. Achieving comparable efficiency with a smaller-volume, higher-DCR inductor underscores the D-SCLB topology’s key advantage: by reducing the average inductor current, it reduces conduction losses.
Fig.~\ref{fig.eff_prof_comp_vin} (a) and (b) confirm the accuracy of the model\textcolor{black}{()} prediction with simulation and measurement results for D-SCLB and SCLB converters, respectively under a constant load resistance of 2.5$\Omega$ and the input voltage is varied from 40\,V to 60\,V. The theoretical model predictions are within 1.5\% for both the converters. 
Fig.~\ref{fig.eff_prof_comp_vin} (c) and (d) confirm the accuracy of the model prediction under a constant load resistance of 25$\Omega$. For the 25$\Omega$ light load condition,  the theoretical model predictions for the D-SCLB (SCLB) converter are within 1.5\% (2.1\%) of the simulation results, the accuracy of matching to the measurement results is 5.5\% (2\%). The higher deviation of the model prediction is because the switching losses, namely overlap losses, deadtime losses, and diode recovery losses are not accounted for in the model and their impact is higher at lighter loads.

Fig.~\ref{fig.eff_prof_comp_vin}(a) and (b) confirm the model prediction against simulation and measurement results for the D-SCLB and SCLB converters, respectively, under a constant load resistance of 2.5\,$\Omega$ as the input voltage varies from 40\,V to 60\,V. The theoretical model predictions remain within 1.5\% for both converters. Fig.~\ref{fig.eff_prof_comp_vin}(c) and (d) confirm the model prediction under a constant load resistance of 25\,$\Omega$. For the 25\,$\Omega$ light-load condition, the theoretical model predictions for the D-SCLB and SCLB converters are within 1.5\% and 2.1\%, respectively, of the simulation results; the accuracy of matching to the measurement results is 5.5\% for D-SCLB and 2\% for SCLB. The higher deviation at light load arises because switching losses, namely overlap losses, dead-time losses, and diode-recovery losses, are not accounted for in the model, and their impact is greater under light-load conditions.

\section{Conclusion}
\label{sec.conc}
This paper has presented the D-SCLB, an inductively assisted $N$:1 SC ladder buck converter that replaces the output-side switch in the bottom half-bridge with a small inductor to lower the average inductor current to $(1{-}M)I_{out}$ while preserving the ladder's optimal switch $V$-$A$ stress. Closed-form output-impedance models, derived for both optimal and uniform switch sizing, show that the 4:1 D-SCLB improves $R_{out}$ by up to 30\% and reduces the average inductor current by up to 20\% compared to the SCLB. Two hardware prototypes on a low-cost two-layer PCB confirm a peak efficiency of 96.2\% for the D-SCLB, which uses a 49\% smaller inductor volume than the SCLB at comparable efficiency, with model predictions within 1.5\% of measurements under heavy load. The topology is directly extensible to higher conversion ratios and is amenable to monolithic integration.

\vspace{12pt}

\end{document}